\documentclass{ws-rv9x6}
\usepackage{subfigure}   % required only when side-by-side / subfigures are used
\usepackage{ws-rv-thm}   % comment this line when `amsthm / theorem / ntheorem` package is used
\usepackage{ws-rv-van}   % numbered citation & references (default)
\makeindex
\crop[off]

\makeatletter
\renewcommand\@makechapterhead[1]{%
    {\vbox to 110pt{%
    \def\thefootnote{\@fnsymbol\c@footnote}%
    \vspace*{33pt}%
        \parindent\z@\raggedright\reset@font
        {\centering{{\normalfont\fontsize{10}{12}\selectfont
            Contribution to the edited volume\\[2pt]
            \textit{Recent Progress in Computational String Geometry}\par}%
         \vskip 0.25in
    \vbox{\CTfont #1\par}\par}\par}\nobreak\vfill}}}%
\makeatother

\begin{document}

\chapter[Generating Special Triangulations with Transformers]{Generating Special Triangulations with Transformers\label{ra_ch1}}

\author[C. Arnal, J.H.T. Yip, F. Charton, and G. Shiu]{Charles Arnal,\footnote{Equal contribution, randomized order.}$^,$\footnote{charles.a.arnal@gmail.com} Jacky H. T. Yip,$^{*,}$\footnote{hyip2@wisc.edu} François Charton,\footnote{fcharton@gmail.com} and Gary Shiu\footnote{shiu@physics.wisc.edu}}
%\index[aindx]{Author, F.} % or \aindx{Author, F.}
%\index[aindx]{Author, S.} % or \aindx{Author, S.}

\address{$^\dagger$FAIR at Meta, \\
6 Rue Ménars, 75002 Paris, France}
\address{$^{\ddagger,\P}$Department of Physics, University of Wisconsin–Madison, \\Madison, Wisconsin 53706, USA}
\address{$^\S$Axiom Math, \\
124 University Avenue, Palo Alto, California 94301, USA}

\begin{abstract}
%This chapter elaborates on the presentation given by G.~Shiu %at the Chennai Mathematical Institute Workshop: \emph{Recent %Progress in Computational String Geometry} in January 2026. %Further physical and mathematical details can be found in the %original paper.\cite{Yip:2025hon}\\\\
%This chapter elaborates on the presentation given by G.~Shiu %at the Chennai Mathematical Institute Workshop: \emph{Recent %Progress in Computational String Geometry} in January 2026. 
%The emphasis is on the machine learning aspects of  %Ref.!\cite{Yip:2025hon} with a different focus where the %original physical and mathematical results were presented. %This chapter thus serves as supplementary material for a %readership interested in AI for Mathematics and Theoretical %Physics. \\ \\
This chapter expands on the presentation delivered by G.~Shiu at the Chennai Mathematical Institute Workshop, \emph{Recent Progress in Computational String Geometry}, held in January 2026.
The focus here is on the machine learning aspects of Ref.~\cite{Yip:2025hon}, offering a complementary perspective to the original work, where the primary physical and mathematical results were presented. As such, this chapter is intended as supplementary material for readers interested in applications of artificial intelligence in mathematics and theoretical physics. \\ \\

Triangulations, i.e., well-structured decompositions of geometric objects into triangle-like pieces, are central objects in many domains of mathematics and physics.
In particular, fine, regular, and star triangulations (FRSTs) of $4$D reflexive polytopes give rise to smooth Calabi-Yau threefolds, which are of significant interest in string theory. 
However, the high dimensionality and combinatorial complexity of triangulations make them particularly challenging to model with classical numerical methods or machine learning.
In this work, we show that transformers, equipped with an appropriate encoding scheme, can be effectively trained to representatively generate new FRSTs across a range of polytope sizes. Moreover, these models can also self-improve through retraining on their own output. This opens the door to both concrete applications to the classification of Calabi-Yau manifolds and further research in physics, combinatorics and algebraic geometry.
\end{abstract}

\markboth{C. Arnal, J. H. T. Yip, F. Charton \& G. Shiu}{Generating Special Triangulations with Transformers}% Customized running heads

\body

%\tableofcontents

\section{Introduction}
While machine learning has been applied with great success to the study of graphs \cite{kipf2016semisupervised,hamilton2017inductive,velikovi2017graph,Perozzi_2014,grover2016node2vec,gilmer2017neural,zhou2018graph,Wu_2021}, higher dimensional geometric structures remain underexplored.
In particular, \textit{triangulations}--well-structured decompositions of geometric objects into triangle-like pieces called \textit{simplices}--have scarcely been studied \cite{sharp2020pointtrinet,lei2023circnet}, despite their key role in several domains of mathematics and physics, such as combinatorics \cite{combinatorics1, Neumann+1992+243+272,  Lackenby2000, Kristof}, algebraic geometry \cite{Ragsdale, viro2006patchworkingrealalgebraicvarieties, Arnal_2022}, topological data analysis \cite{Chazal2017AnIT, Carlsson_Vejdemo-Johansson_2021} and string theory \cite{Batyrev:1993oya,Kreuzer_2001,Altman:2014bfa}.
Human researchers often struggle to visualize and manipulate higher dimensional triangulations due to the limitations of human visual intuition; consequently, many problems which require defining special triangulations with desirable properties remain open.
Machine learning methods do not face the same limitations and might prove successful at modelling interesting distributions of triangulations, as has been the case for other classes of combinatorial objects \cite{wagner2021constructionscombinatoricsneuralnetworks, novikov2025alphaevolvecodingagentscientific}.

String theory is a particularly promising area of application.
Understanding the string landscape--determining string theories that are compatible with realistic particle physics and gravity--is a challenging task because of the number of possible choices for compactification: packing the many dimensions of string theory into the four dimensions of physical space. 
Fine, regular, star triangulations (FRSTs) of $4$D reflexive polytopes give rise to toric Calabi-Yau threefolds, which in turn provide $4$D vacuum configurations of superstring theory that can accommodate realistic physics~\cite{Candelas:1985en,Marchesano:2024gul}.
As the number of $4$D reflexive polytopes is large but finite--$473{,}800{,}776$ in total~\cite{Kreuzer:2000xy}--all possible configurations could theoretically be enumerated by considering all the possible triangulations of each of these polytopes.
The number of FRSTs grows exponentially with the number of vertices in the polytope, which makes this strategy computationally prohibitive.
A more reasonable strategy is to develop efficient techniques for generating diverse FRSTs for large polytopes.

In this paper, we propose a generative technique for FRSTs based on encoder-decoder transformers~\cite{Vaswani:2017lxt}, which we call \textit{CYTransformer}.
Our models are trained on known FRSTs, encoded as sequences of tokens representing simplices, to predict triangulations of input polytopes encoded as sequences of coordinates describing their vertices. 
In greater detail, our key contributions are as follows:
\begin{itemize}
    \item We propose a new \textit{transformer-compatible encoding of polytopes and triangulations}, along with an adapted architecture and domain-specific \textit{data augmentation techniques} that prove to be crucial.\\
    
    \item We show that once trained on a particular set of polytopes and associated FRSTs, CYTransformers are not only capable of generating new FRSTs, but can also \textit{generalize to yet unseen polytopes}.\\
    
    \item We show that CYTransformers' output distribution is highly \textit{representative} of the full ensemble of FRSTs.\\
    
    \item We explore \textit{self-improvement techniques} and observe that CYTransformers can continuously improve beyond their initial training data by repeatedly generating new FRSTs and retraining on them--a form of reinforcement learning with rejection sampling.
\end{itemize}
We also conduct ablations of the relevant hyperparameters. 
Those results, and in particular the very fact that transformers are capable of learning the distribution of FRSTs, are striking given the combinatorial complexity of that distribution and how different it is from the natural-language data on which transformers are already known to excel.
While this article focuses on the specific case of FRSTs, these findings open the door to future research on the learning of other higher-order geometric structures of interest to mathematicians and physicists, such as hypergraphs. 

\section{Related Works}
Current FRST-generating approaches comprise non-learning algorithms involving bistellar flips~\cite{10.5555/1952022}, random walks, or both, which tend not to scale well with the number of vertices of the polytopes considered~\cite{Demirtas:2020dbm}. 
On the other hand, genetic \cite{MacFadden:2024him} and reinforcement learning \cite{Berglund:2024reu} techniques do not easily transfer from one polytope to another, and the representativeness of their output distributions has not been demonstrated.
These limitations can be attributed to the combinatorially explosive nature of the space of triangulations, as well as to the difficulty in finding an amenable encoding of triangulations.

We hope to overcome these difficulties thanks to the flexibility and scalability of transformers.
While transformers' ability to model natural language and programming code is now well-established~\cite{Touvron2023LLaMAOA,Jiang2023Mistral7,Achiam2023GPT4TR,Llama3}, they have also been applied to more exotic data distributions coming from mathematics~\cite{Funsearch, charton2024patternboost, Alfarano, hashemi2025transformersenumerativegeometry}, physics~\cite{Geneva2020TransformersFM, Janny2023EagleLL,  Cai:2024znx, Smatrix}, or chemistry~\cite{jumper2021highly}. Unlike natural and programming languages, which have shared grammatical and logical structures~\cite{Inductionheads, Iterationhead, FunctionVI} that transformers are known to be well-suited for, data distributions from new scientific problems vary widely, creating unique challenges. Finding efficient and transformer-compatible representations for such problems is a non-trivial task.

\section{Toric Calabi-Yau manifolds and Fine Regular Star Triangulations (FRSTs)}
We give the necessary theoretical background, with a focus on the elements needed to understand the machine learning pipeline presented in the next section. %We refer to Appendix~\ref{app:CY_theory} for additional details.

\subsection{Triangulations}
\begin{figure}
% \vspace{-10pt}
    \centering
    \includegraphics[width=0.25\textwidth]{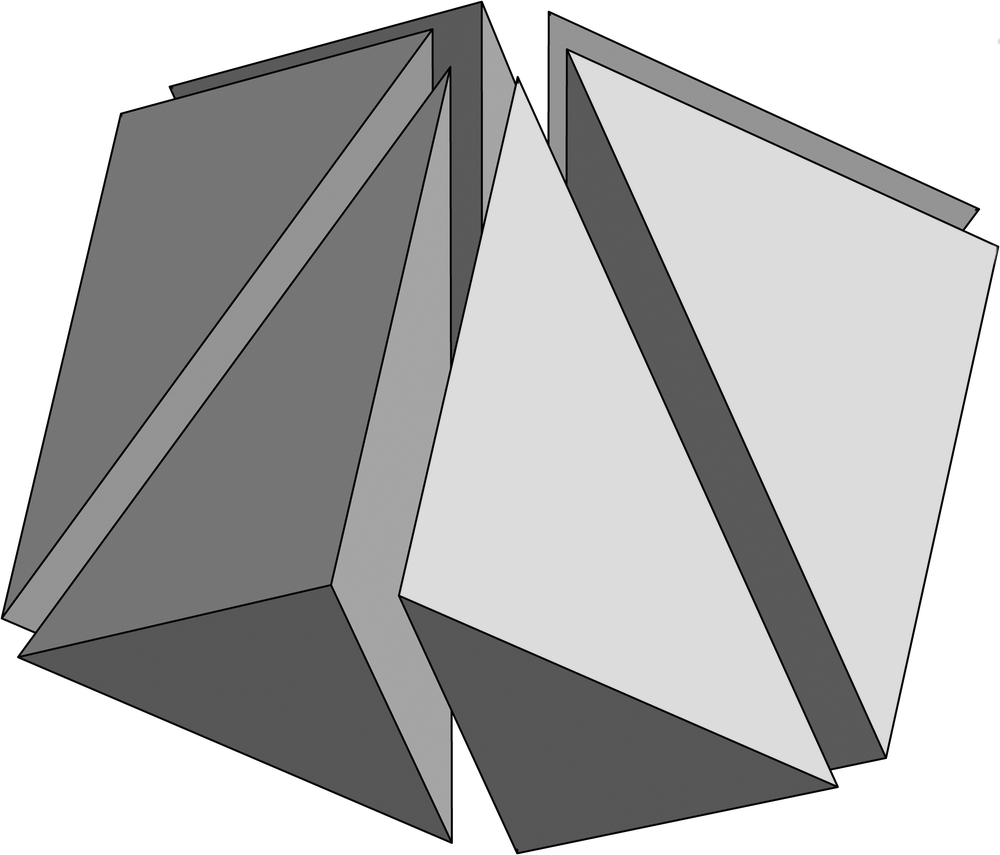}
    \caption{A triangulation of a $3$D cube.\cite{Kerber2016ConstrainedTV}}
    \label{fig:example_triangulation}
    \vspace{-10pt}
\end{figure}

A $d$-dimensional \textit{simplex} in $\mathbb{R} ^ n$ is the convex hull of $d+1$ affinely independent points in $\mathbb{R}^n$--segments and triangles  are examples of $1$D and $2$D simplices respectively.  
Given an $n$-dimensional polytope $\Delta \subset \mathbb{R} ^ n$ (i.e., the convex hull of finitely many points that affinely span $\mathbb{R}^n$ ), a \textit{triangulation} of $\Delta$ is a set of $n$-dimensional simplices $S_1, \ldots, S_k \subset \Delta$ such that their union is $\Delta$, and such that any two distinct simplices $S_i, S_j$ either do not intersect or intersect along a common face.
One case of particular interest is when the triangulation has to be supported on the vertices of the polytope, i.e., when the vertices of any of its simplices have to be vertices of $\Delta$ (as opposed to interior points).
An illustration is given in Figure~\ref{fig:example_triangulation}.

Triangulations are among the most frequent higher-dimensional geometric and combinatorial objects in mathematics, be it in topology, geometry, algebraic geometry or combinatorics. In some regards, they can be seen as higher-dimensional analogues of graphs.
Despite their ubiquitousness, they are difficult to handle by either human intuition or automated methods. This is in part due to the number of possible triangulations of a polytope typically growing exponentially fast in the number of its vertices.
Furthermore, while triangulations are very numerous, they are also very rare among all the possible collections of sub-simplices of a given polytope, due to the rigid union and intersection conditions given above, which often makes them impossible to enumerate using brute force methods.

\subsection{Calabi-Yau manifolds, reflexive polytopes, and FRSTs}
As mentioned in the introduction, Calabi-Yau manifolds play a crucial role in string theory. They are smooth and compact spaces equipped with a Ricci-flat Kähler metric, and are prime candidates for string compactification: the Ricci-flatness preserves supersymmetry, while the toric structure endows these spaces with torus symmetries needed in particle physics model building \cite{Candelas:1985en,Katz_1997}.

One of the ways to define toric Calabi-Yau manifolds is with 
\textit{reflexive polytopes} and their \textit{fine, regular, and star triangulations (FRSTs)}.
A reflexive polytope $\Delta$ is the convex hull of a finite set of points of $\mathbb{Z}^n$ such that its dual polytope $\Delta^\circ := {\text{Conv}}(\{y\in \mathbb{Z}^n\mid\langle m,y\rangle\geq -1\;\forall m\in\Delta\})$
is also the convex hull of a finite set of points of $\mathbb{Z}^n$.
It can be shown that the interior of a reflexive polytope contains exactly one point of $\mathbb{Z}^n$, which is taken to be the origin.
The set of points consisting of all points in $\Delta \cap \mathbb{Z}^n$, excluding those strictly interior to codimension-1 faces, together with the origin, is referred to as the set of resolved vertices of $\Delta$.
Up to lattice automorphisms in $GL(n,\mathbb{Z})$, the number of reflexive polytopes in any given dimension $n$ is finite.
In the remainder of this paper, we split and organize reflexive polytopes according to the number $N_{\text{vert}} $ of their vertices.\footnote{One could also consider the $(1,1)$-Hodge number $h^{1,1}$ of the corresponding Calabi-Yau manifolds, which is of greater interest to physicists. To simplify the presentation of our work, we restrict ourselves to polytopes such that $N_{\text{vert}}=h^{1,1}+5$, which is the case for most polytopes with a small number of vertices. }

Given a reflexive polytope, we are interested in special triangulations of its resolved vertices (i.e., triangulations such that the vertices of their simplices are resolved vertices). Those triangulations $\mathcal{T}$ must satisfy: every resolved vertex appears as a vertex of some simplex in $\mathcal{T}$  (\textbf{Fine (F)}), $\mathcal{T}$ arises as the projection of the lower faces of a convex polytope constructed from lifting each resolved vertex $p_i$ to $(p_i,h_i)\in \mathbb{Z}^n\times\mathbb{R}$ with some height $h_i\in\mathbb{R}$ (\textbf{Regular (R)}), and every simplex in $\mathcal{T}$ contains the origin as a vertex (\textbf{Star (S)}).
Hence we call these FRSTs.
Given a reflexive polytope $\Delta$ and an FRST $\mathcal{T}$ of $\Delta$, a classical algebraic construction yields a smooth, projective, and compact Calabi-Yau threefold, which is of special interest in string theory \cite{Batyrev:1993oya,Demirtas:2020dbm}.

\section{Generating special triangulations with
transformers}\label{sec:method}
Our goal is to train a model that takes as input a reflexive polytope and outputs triangulations of interest of that polytope--in our case study, FRSTs.
A key difficulty lies in finding amenable encodings of polytopes, which are sets of points, and of triangulations, which are sets of sets of points, as well as powerful model architectures compatible with those encodings.
The encodings must in particular be flexible enough to allow the transfer of knowledge from one polytope to another.
We call our proposed solution \textbf{CYTransformer}.

\subsection{Encoding polytopes and triangulations}

We encode four-dimensional reflexive polytopes as sequences of $4$D vectors representing their vertices (with the origin omitted).
Collections of simplices (triangulations or not) are represented as sequences of 4-simplices containing the origin. For polytopes of fixed size $N_{\rm vert}$, there are ${N_{\rm vert}-1\choose 4}$ possible such simplices, which we encode as as many discrete tokens (with \texttt{<i>} standing for the $i$-th element in the enumeration of possible 4-simplices). In other words, we fix a mapping $F$ between the subsets of cardinality $4$ of $\{1,\ldots, N_{\text{vert}} -1\}$ and the integers $\{1,\ldots, {N_\text{vert}-1\choose 4}\}$ (e.g. using the lexicographic order).
For a given input polytope $\Delta$ with vertices $(V_1, \ldots, V_{N_{\text{vert}} -1})$, we then represent a star simplex $\{\text{origin},V_{i_1},V_{i_2},V_{i_3},V_{i_4}\}$ by the image by $F$ of the set $\{i_1,i_2,i_3,i_4\}$.
As a consequence, the meaning of each of these tokens is always conditioned not only on the input polytope, but also on the order in which the vertices of the input polytope appear in its encoding.
We also use special begin-of-sequence, end-of-sequence, and padding tokens. %Concrete examples of encodings are given in Appendix~\ref{app:encodings}.

\subsection{Model architecture}

\begin{figure*}[t]
    \centering
    % \resizebox{0.85\textwidth}{!}{\input{plots/full_architecture}}
    % % \vspace{-1em}
    \includegraphics[width=0.85\textwidth]{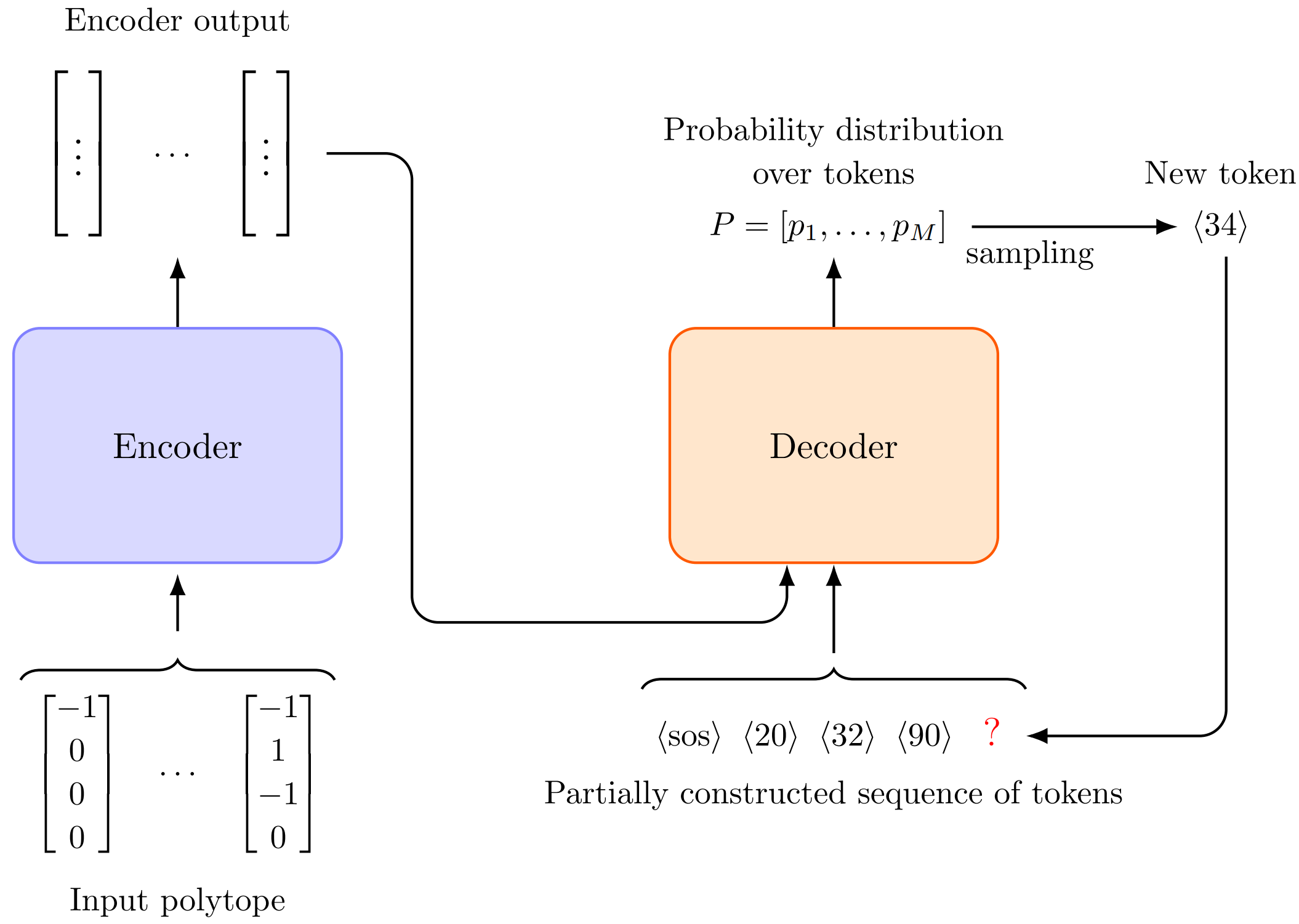}
    \caption{\textbf{CYTransformer architecture.} The high-level pipeline for our model in inference mode. The encoder processes the input polytope, as a sequence of four-dimensional vertex vectors, into a latent representation. The decoder autoregressively generates tokens, representing simplices, conditioned on both the encoder output and previously generated tokens.
    }
    \label{fig:full_architecture}
    % \vspace{-1em}
\end{figure*}

We adopt the encoder-decoder transformer architecture introduced in the landmark paper~\cite{Vaswani:2017lxt}.$^,$\footnote{We believe the encoder-decoder architecture is well-suited to our problem, because a separate bidirectional encoder can learn a richer representation of input polytopes, and can be cached at inference for faster generation.}
The high-level architecture, which we also illustrate in Figure~\ref{fig:full_architecture}, is as follows: at any given step, the model takes as input the entire input polytope and a partially constructed sequence of tokens (the candidate triangulation in construction). Each token represents a simplex. The input polytope is processed by the encoder to produce a sequence of high-dimensional vectors: the encoder output. The encoder output, together with the sequence of tokens generated so far, are used as inputs by the decoder. The decoder then outputs a vector $\boldsymbol{P}$ representing a probability distribution over the set of all tokens considered.
At inference time, a new token is sampled from this probability distribution, and added to the sequence being constructed. This process is repeated until the special end of sequence token is produced, or until the sequence reaches a predefined length.

We train the model on training sets of pairs $(P,T)$, where $P$ is a reflexive polytope (with a fixed number $N_{\rm vert}$ of resolved vertices) and $T$ is an FRST of $P$.
For a given pair of polytope and FRST, CYTransformer is trained to minimize the cross-entropy loss between the predicted sequence (the triangulation it predicts for this polytope) and the FRST from the training set. Seen from a language model perspective, we proceed as if the polytope's encoding was a prompt and the associated FRST was a correct answer.

Note that since the model only learns from examples, it has no a priori knowledge of what an FRST is. Therefore, there is no guarantee that the output of a trained CYTransformer describes a valid triangulation, let alone an FRST.
Note also that while we choose to apply this training pipeline to the specific case of reflexive polytopes and FRSTs, it could a priori be used with any distribution of polytopes and associated triangulations.

\begin{figure*}[t]
    \centering
  \makebox[\textwidth][c]{%
  \includegraphics[height=4cm]{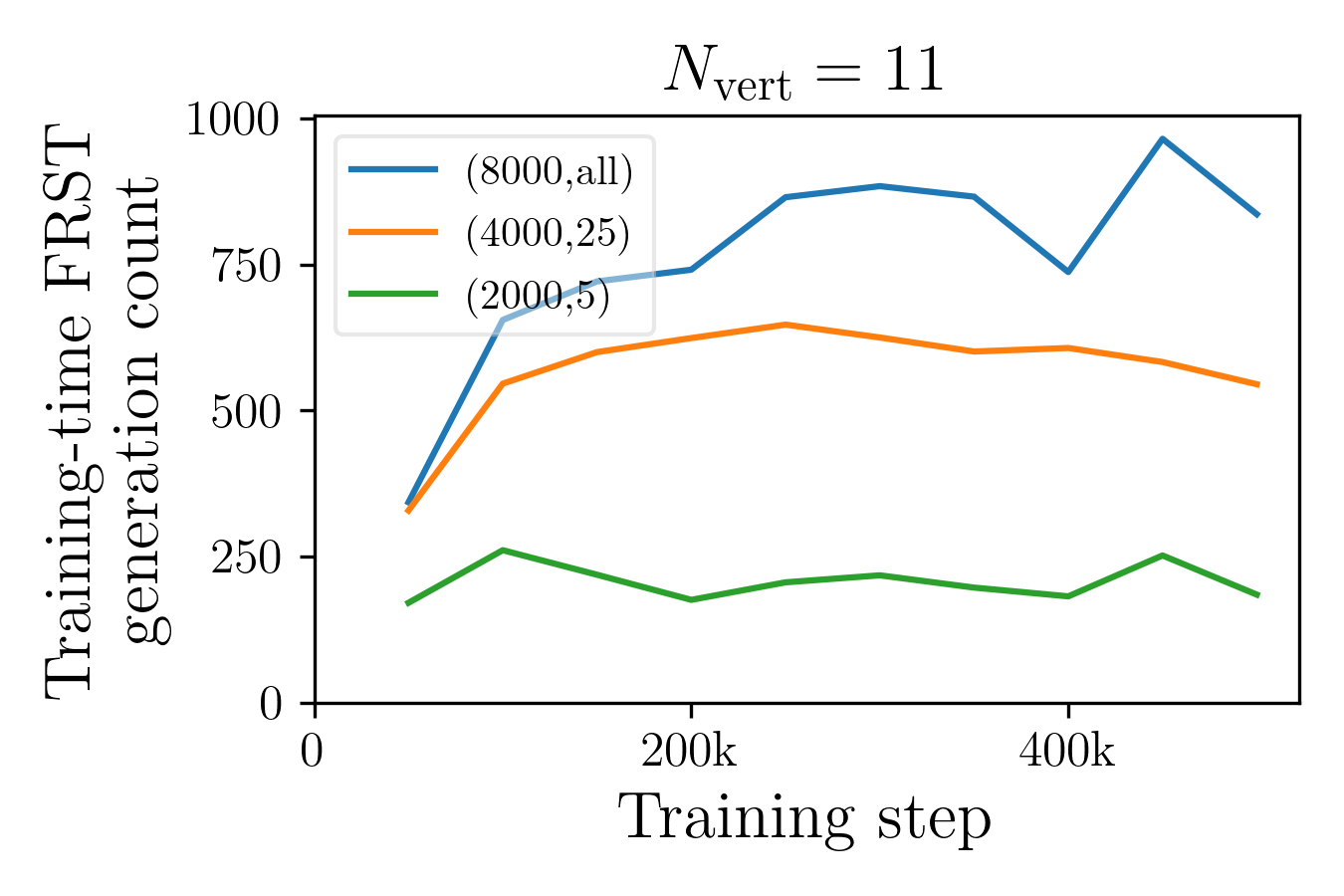}
  \includegraphics[height=4cm,trim={1.0cm 0 0 0},clip]{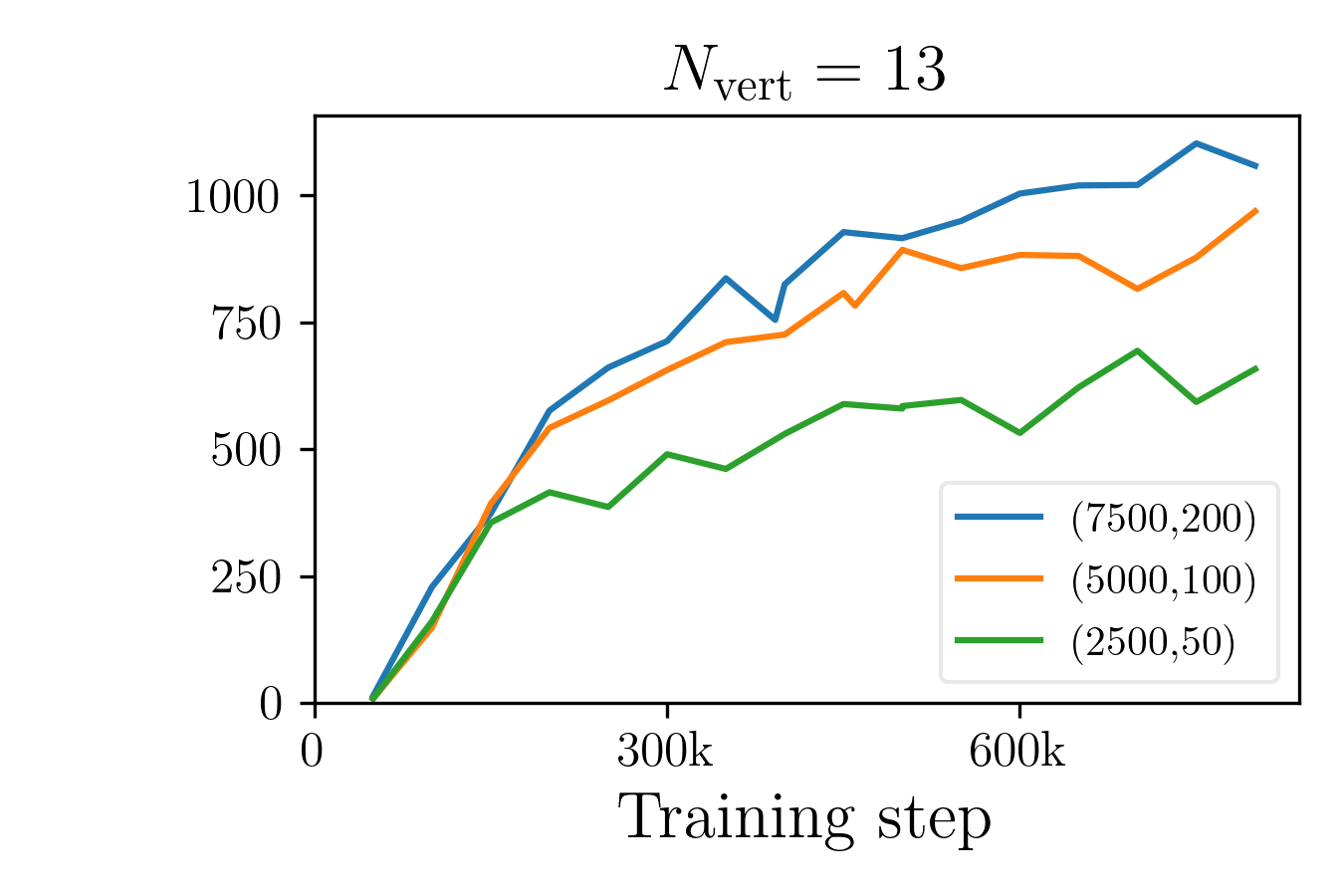}}
  \includegraphics[height=4cm,trim={1.0cm 0 0 0},clip]{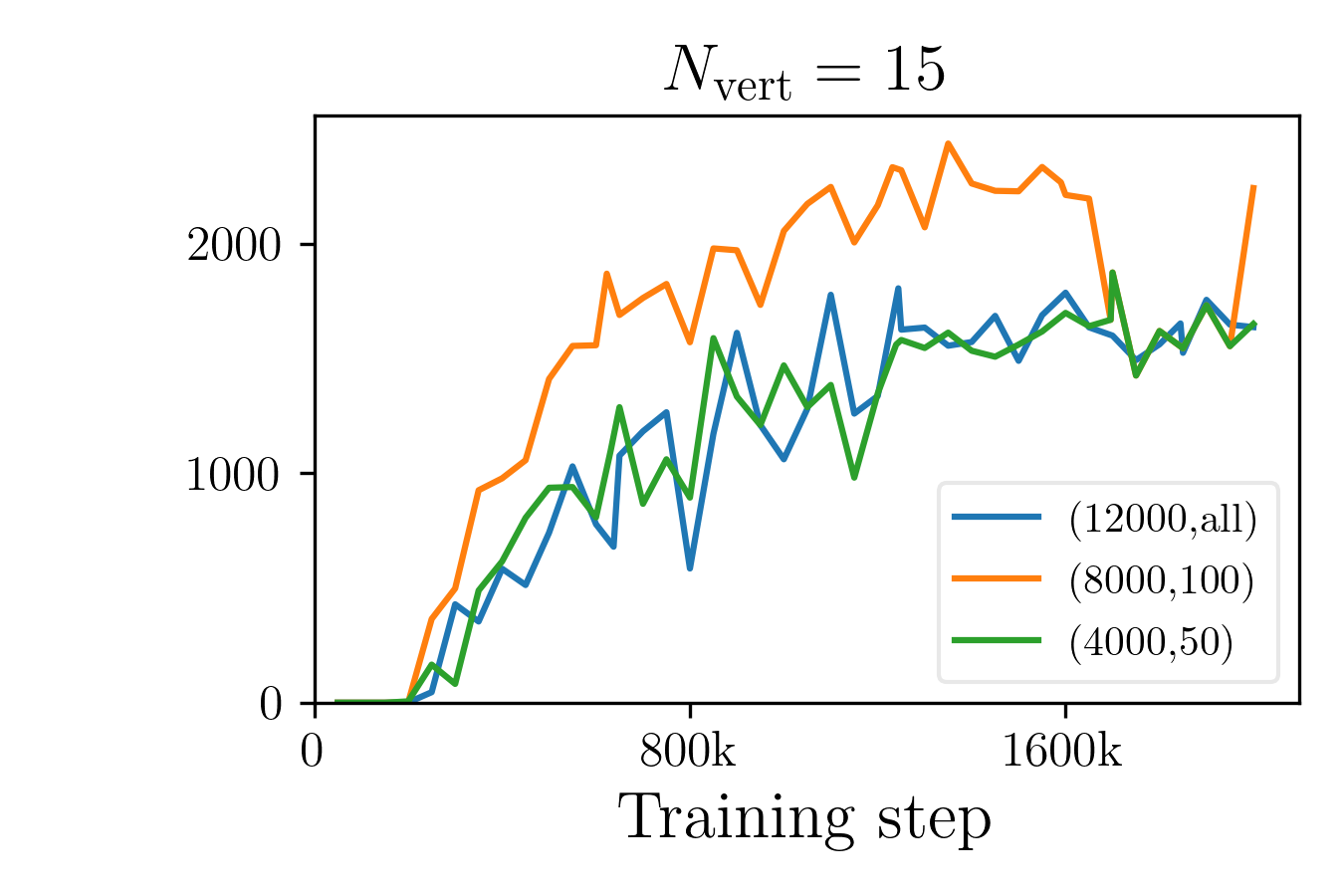}
  \caption{ \textbf{Training dynamics for various training set sizes.} 
  FRST generation count measured across $1{,}600$ (for $N_{\rm vert}=11,13$) or $6{,}400$ (for $15$) candidate triangulations as a function of training step.  Each curve within a plot corresponds to a model trained on a different dataset size $(N_{\rm polys}, N_{\rm triangs})$, where $N_{\rm polys}$ is the total number of polytopes and $N_{\rm triangs}$ is the maximum number of FRSTs per polytope (with ``all'' being all available FRSTs). 
  }
  \label{fig:training_dynamics}
\end{figure*}

\subsection{Data augmentation}\label{subsec:data_augmentation}

Polytopes and their associated triangulations are invariant to permutations of the polytopes' vertices. Furthermore, a triangulation is invariant under permutations of the order of its simplices. We want our models to be invariant to these symmetries. To this end, every time a training example is selected, we apply the following data augmentation procedures:
first, the polytope vertices are randomly permuted, and the tokens in the output sequence are replaced accordingly.
Second, the simplices in the output sequence are also randomly permuted.

This guarantees that the trained CYTransformer is invariant to these transformations, which are mere artifacts of our tokenization scheme. Other known symmetries of reflexive polytopes, involving combinations of reflections, rotations, and shears, are not considered.\footnote{Though preliminary experiments suggest they might slightly influence the training dynamics.}

\subsection{Self-improvement}\label{subsec:selfim_method}
The quality of a machine learning model is constrained by the size and diversity of its training data, and this applies to our models (see section~\ref{subsec:training_dynamics}). For small values of $N_{\rm vert}$, classical algorithms can generate all FRSTs of a polytope, or a large enough sample to provide a representative subset (see further below). However, for larger numbers of vertices, this process becomes prohibitively costly. To reduce dependence on this expensive and uncertain data preparation step, we explore a form of self-improvement. The core idea is simple: use the model's own generated FRSTs to augment its training data for retraining, and thus alleviate the data scarcity. While our current experiments are conducted on moderate values of $N_{\rm vert}$, this serves as a proof of concept for a strategy that becomes especially valuable when scaling to larger configurations.

Concretely, we begin by training CYTransformer on a small initial dataset. After a fixed number of training steps, we pause and use the current model to generate candidate triangulations for polytopes drawn from the training set.\footnote{Note that there is no test set contamination here, as this remains part of the training procedure.} Valid FRSTs among the candidates are identified and added to the training set, expanding its size. The model is then retrained for another iteration on this augmented data, and this cycle is repeated until its performance starts stagnating. Since the model incrementally generates more training data for itself, this approach is expected to require only a modest amount of initial data to achieve appreciable performance.

As the model is trained on samples of its own generation, this can be seen as a form of off-policy reinforcement learning, an approach that has become popular for training transformers (and in particular large language models) in recent years \cite{R1, arnal2025asymmetricreinforceoffpolicyreinforcement,  faircodegenteam2025cwmopenweightsllmresearch}. We pursue this reinforcement-learning direction further in~\cite{Yip:2026heterotic}, where a transformer policy is trained to search for line bundle standard models in the heterotic string theory landscape.

\subsection{Training data and validation of candidate FRSTs}
\texttt{CYTools}~\cite{Demirtas:2022hqf} is a recently developed software package, with polytope triangulation as one of its primary design focuses.
It can fetch reflexive polytopes from the complete list of the Kreuzer-Skarke database~\cite{Kreuzer:2000xy}, return FRSTs for a given polytope following various distributions, and check whether a collection of simplices of a given polytope is an FRST.
We use \texttt{CYTools} to generate our initial training data and to check whether the candidate FRSTs generated by our trained models are valid.
We also use its FRST-generating algorithms as a baseline to which we compare our trained models. %See Appendix~\ref{app:CYTools_alg} for details.

\subsection{Experimental setup}\label{subsec:experimental_setup}

All models comprise $16$ encoder and $16$ decoder layers, with an embedding dimension of $512$ and $16$ attention heads in each attention layer. Both the feed-forward network and the encoder input embedding layer are multilayer perceptrons, each with a single hidden layer scaled to four times the embedding dimension. Sinusoidal positional encoding is employed, and all activations are ReLUs. The resulting model contains $\sim120$ million learnable parameters.
%See Appendix~\ref{app:exp_details} for the details of our optimization hyperparameters.

For each choice of polytope size $N_{\rm vert}$ and each training instance, we create three disjoint datasets of polytopes: a training set, a validation set used to measure model performance during training and to decide the stopping point, and a test set reserved for evaluating the trained model.
Depending on the size $N_{\rm vert}$ of the polytopes considered, the training sets contain between $1000$ and $16000$ polytopes, with a maximum of $11$ to $300$ FRSTs per polytope. %Additional details can be found in Appendix~\ref{app:exp_details}.

\begin{figure*}[t]
    \centering
    \makebox[\textwidth][c]{%
  \includegraphics[height=4.7cm,trim={0 0 0 0},clip]{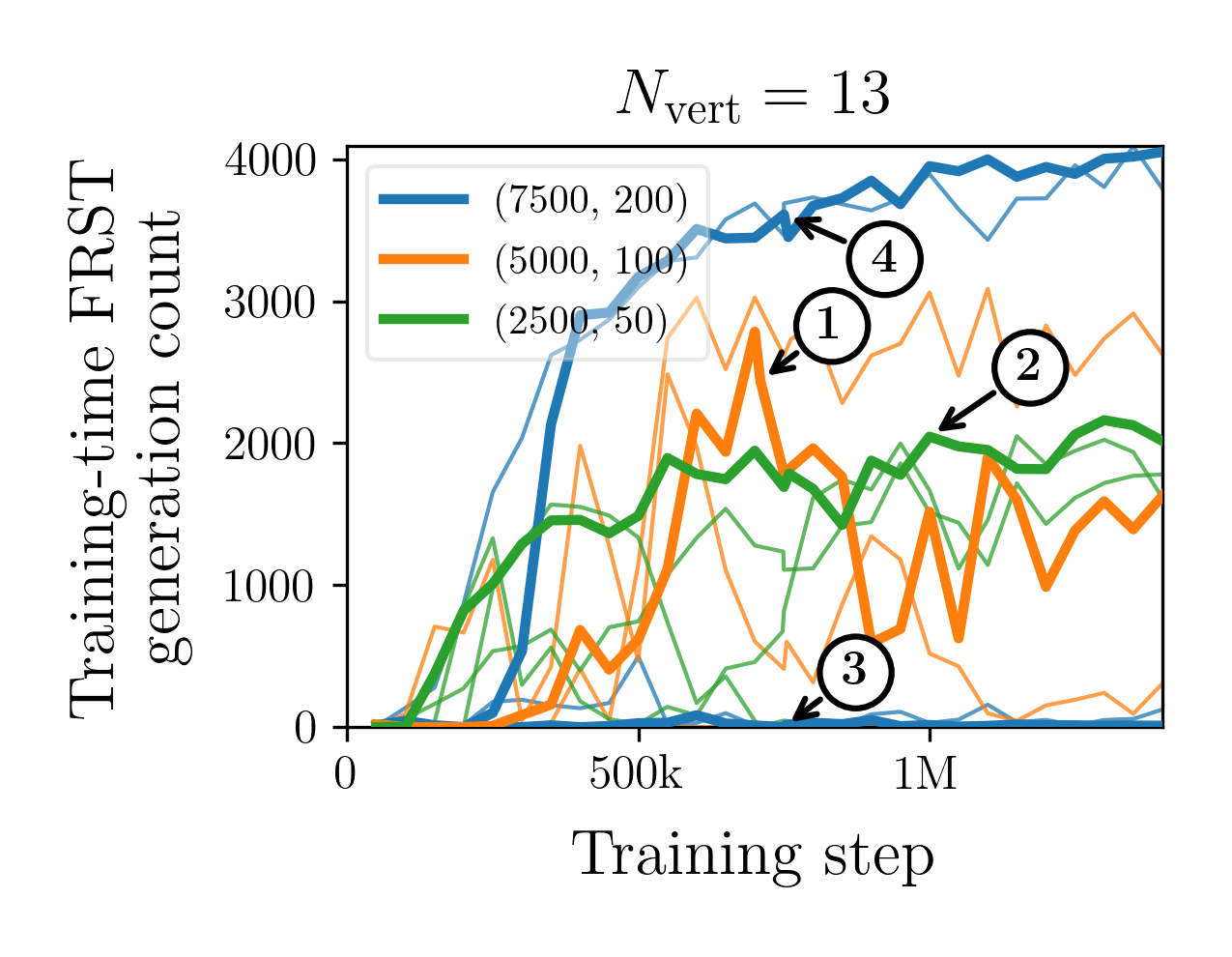}}\\
  \includegraphics[height=4.3cm,trim={0.5cm 0 0 0},clip]{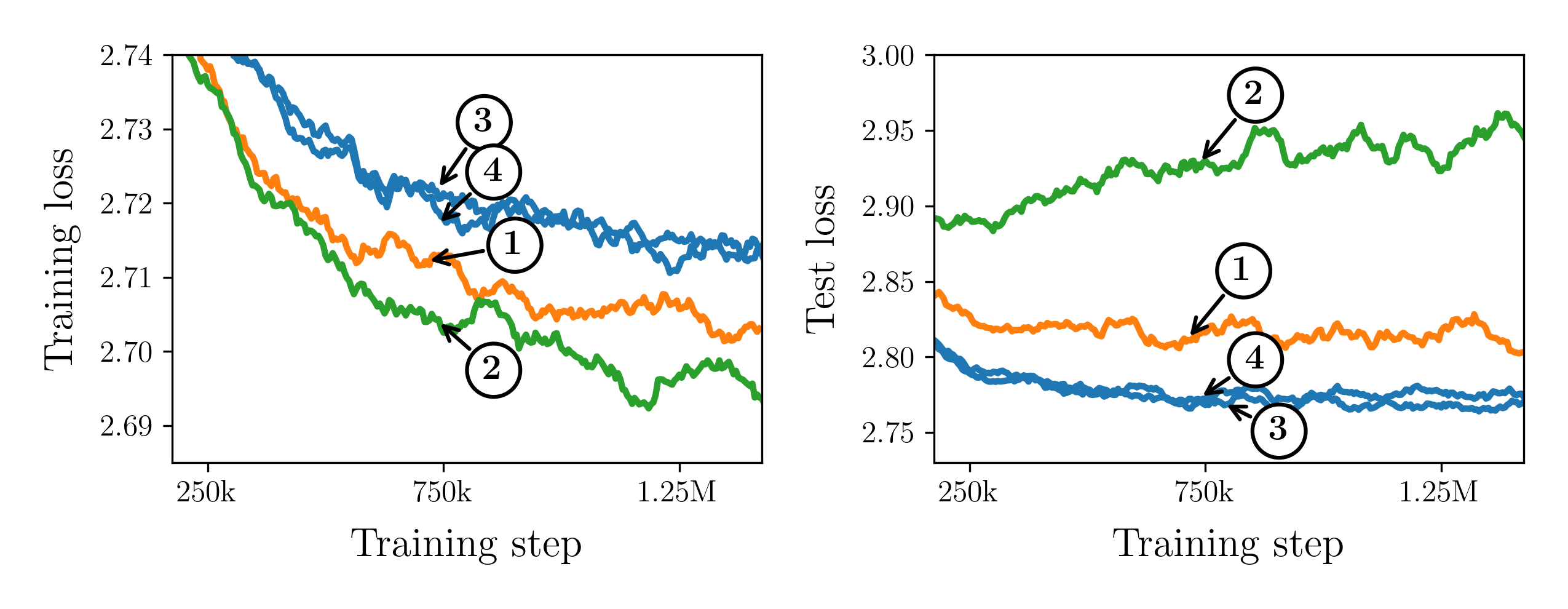}
  \caption{
  \textbf{Detailed training dynamics}
    Left: FRST generation count as a function of training step for several training datasets and random seeds. Middle: Training loss for a subset of those runs.
    Right: Test loss for the same subset of runs.
  }
  \label{fig:training_dynamics_detailed}
\end{figure*}

\section{Experiments}
\label{sec:results}

We analyze the training dynamics and the performances of our model, as well as the effectiveness of the self-improvement method.

\subsection{Training dynamics}\label{subsec:training_dynamics}

We monitor the training of our CYTransformer models using the following performance metric, which we call \textit{FRST generation count}: every few thousand steps, we randomly pick $X$ polytopes from the validation set and have the model generate $Y$ candidate triangulations for each. Among the resulting $X\times Y$ candidates, we count the number of valid and distinct FRSTs.

We train CYTransformers for $N_{\mathrm{vert}} \in \{11,13,15\}$ with training sets containing various numbers of polytopes, and various numbers of triangulations per polytope.
We see in Figure~\ref{fig:training_dynamics}
that the trained models eventually learn to generate FRSTs that were not part of their training set, including for input polytopes on which they were never trained. %(see Figure~\ref{fig:training_dynamics_app} in Appendix~\ref{app:add_results} for similar curves for $N_{\mathrm{vert}} \in \{10,12,14\}$).
This is already remarkable, considering the very strict set of conditions that defines FRSTs and their scarcity among sequences of simplices; note that it would typically take hours for a human expert to find an FRST of a small reflexive polytope without using specialized software. 
Interestingly, we observed (not shown in the figure) that the models almost never output star triangulations, fine star triangulations and regular star triangulations; in other words, they learn to generate FRSTs directly, rather than progressively satisfying finer constraints.

Several phenomena are worth noting.
As expected, the learning task becomes more challenging as  $N_{\rm vert}$ grows larger.
We also see that performance generally improves with a larger training set, though not always (e.g. for $N_{\rm vert} = 15$), which could be linked to the results of \cite{charton2024emergentpropertiesrepeatedexamples}.
While we picked stable training trajectories for Figure~\ref{fig:training_dynamics}, we observed that the training tends to be quite unstable, particularly for large $N_{\rm vert}$, and this despite our large batch size and small learning rate relative to the number of parameters.
This can be observed in Figure~\ref{fig:training_dynamics_detailed}, where several seeds are represented, some of which entirely fail to learn.
Curiously, larger training sets seem to increase both the performance of the most successful seeds and the instability of the training.

\newcommand{\tc}[1]{\textcircled{\raisebox{-.9pt}{#1}}}

The interplay between the training loss, the test loss and the FRST generation count is also worth commenting.
One would naively expect that a low test loss would be positively correlated with the FRST generation count, and that a decreasing training loss and increasing test loss would signal overfitting, which would result in a degradation of the generation accuracy.
While this was often the case, we also frequently observed phenomena that go against this conventional wisdom. Several such situations are represented in Figure~\ref{fig:training_dynamics_detailed}: we see that curve \tc{1} enjoys a dramatic increase in generation accuracy at around 500k steps without it being reflected in the losses, that curve \tc{2} suffers from apparent overfitting (increasing test loss, diminishing training loss) without decline in accuracy, and that curve \tc{3} achieves a negligible FRST generation count despite its test loss being comparable to that of the best performing curve \tc{4}.
These examples, which we cannot completely explain yet, showcase the richness of the learning task studied and of its interaction with the transformer architecture.

\paragraph{Data augmentation and model size ablations}

As explained in Section~\ref{sec:method}, we permute both the vertices of the training polytopes and the simplices of the training FRSTs.
Figure~\ref{fig:data_augmentation} in Appendix~\ref{app:add_results} shows that both these data augmentation techniques are crucial, as omitting either dramatically hurts the learning.

While we do not claim optimality, our ablations in Figure~\ref{fig:data_augmentation} in Appendix~\ref{app:add_results} also show that our chosen model size achieves a good balance between power (models x5 smaller cannot learn) and convenience (models x4 larger cannot comfortably fit on our GPUs).

\begin{figure*}[t]
    \centering
    % --- Left Minipage: The Vertical Label ---
    % Adjust the width (0.05\textwidth) depending on the aspect ratio of your label image
    \begin{minipage}{0.05\textwidth}
        \centering
        % Rotate text 90 degrees counter-clockwise
        \rotatebox{90}{\small Avg. \% of all FRSTs recovered} 
    \end{minipage}%
    %
    % --- Right Minipage: The Grid of Plots ---
    \begin{minipage}{0.98\textwidth}
        \centering
        % Row 1
        \includegraphics[width=0.32\textwidth,trim={1.6cm 0 0 0},clip]{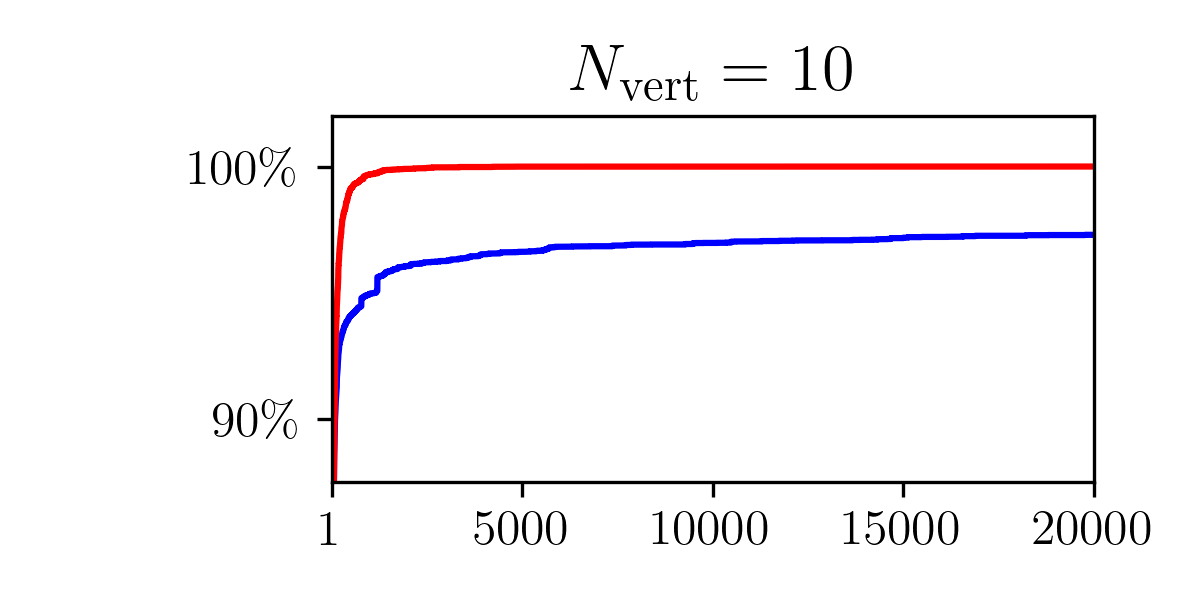}
        \includegraphics[width=0.32\textwidth,trim={1.6cm 0 0 0},clip]{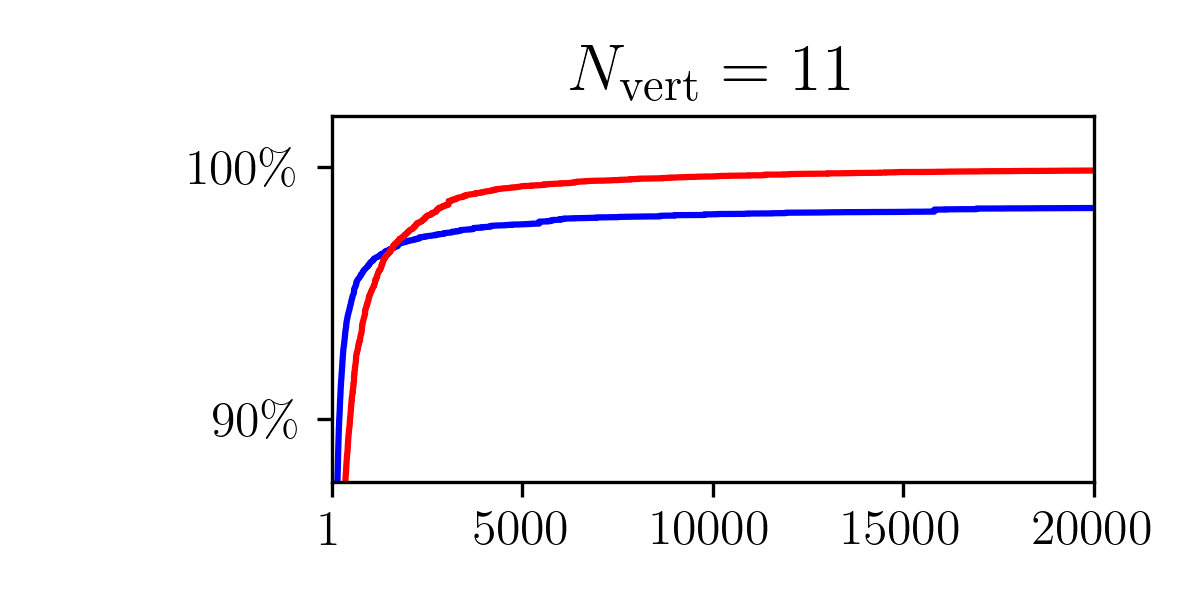}
        \includegraphics[width=0.32\textwidth,trim={1.6cm 0 0 0},clip]{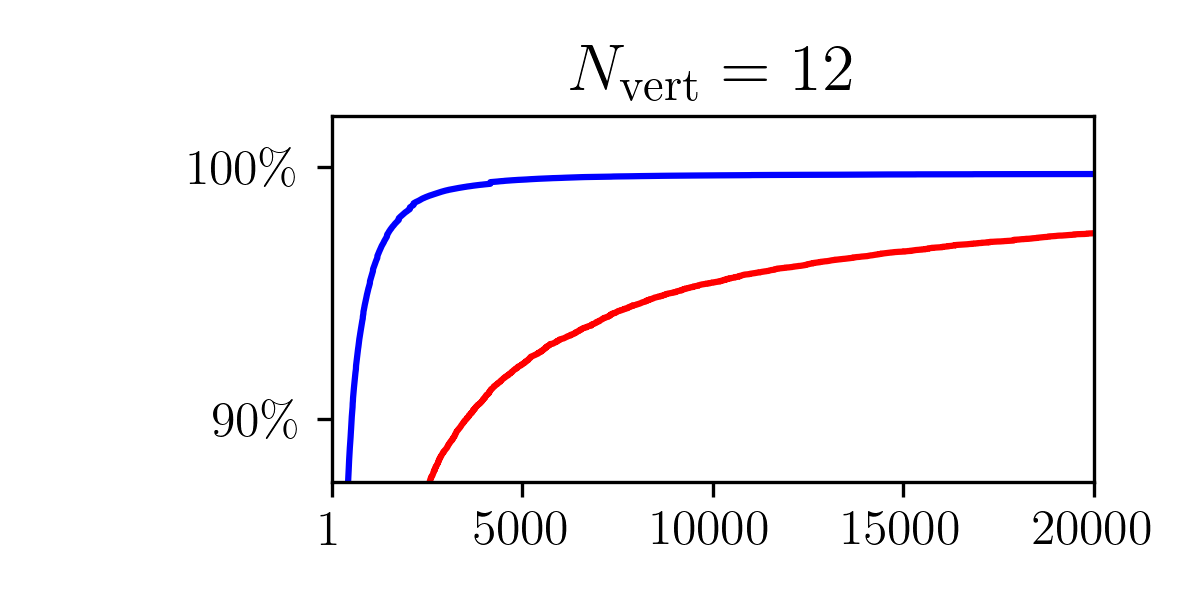} \\
        % \vspace{1mm} % Optional small vertical space between rows
        % Row 2
        \includegraphics[width=0.32\textwidth,trim={1.6cm 0 0 0},clip]{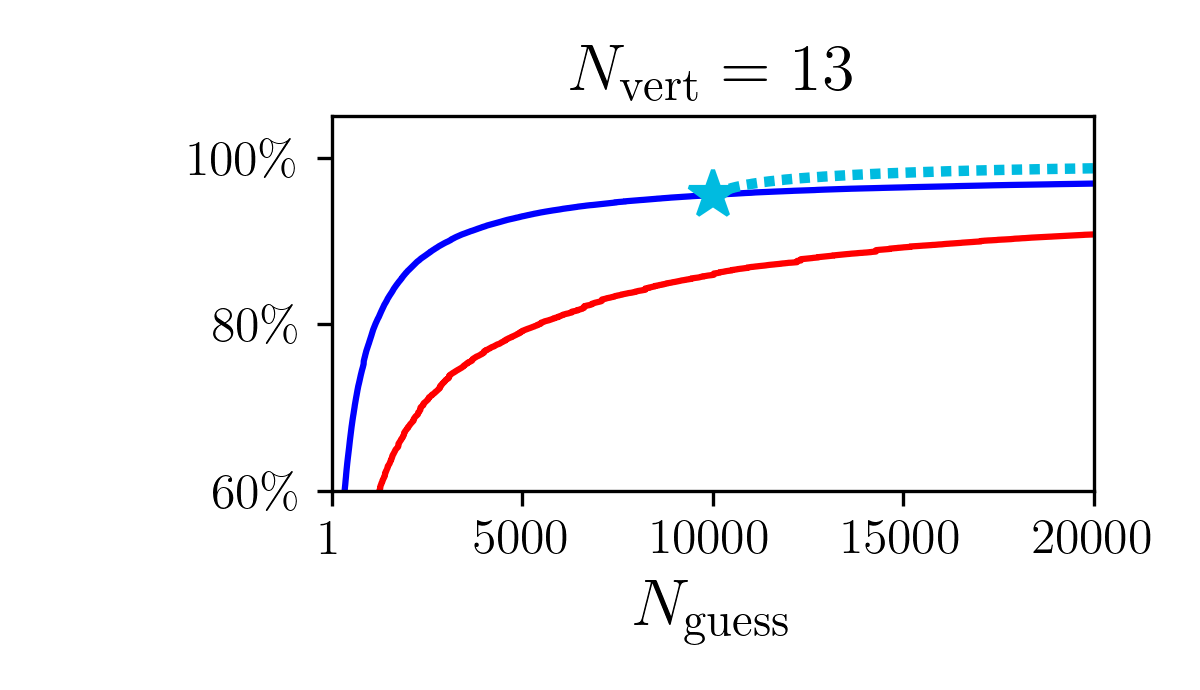}
        \includegraphics[width=0.32\textwidth,trim={1.6cm 0 0 0},clip]{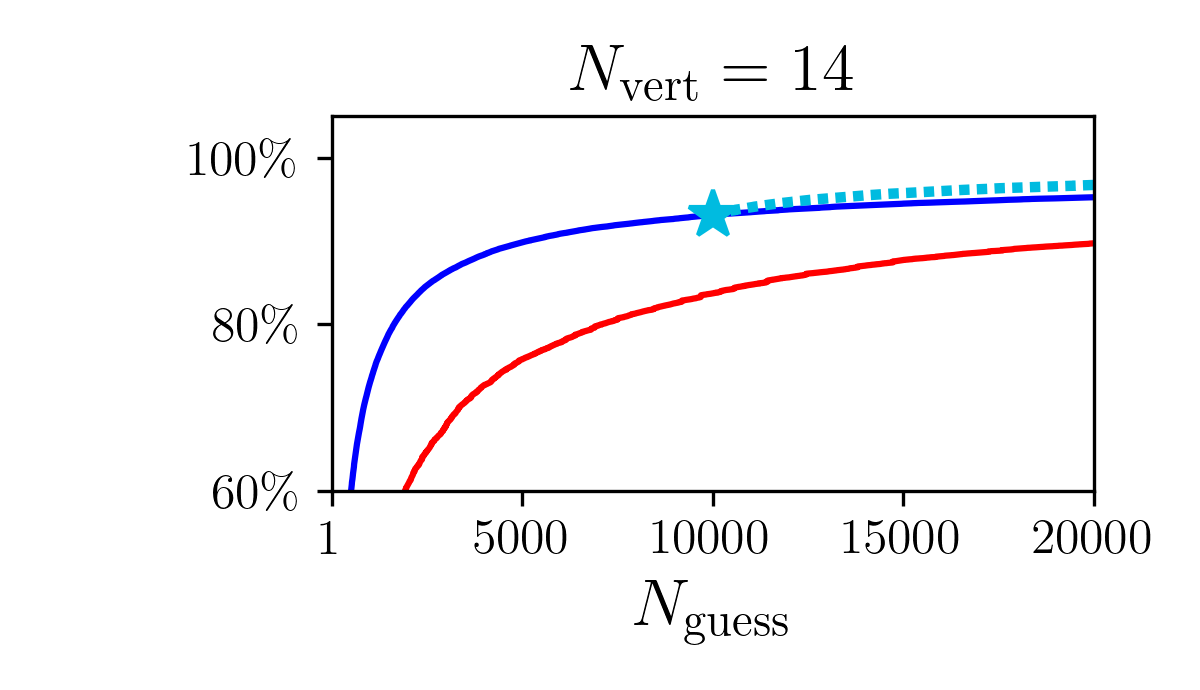}
        \includegraphics[width=0.32\textwidth,trim={1.6cm 0 0 0},clip]{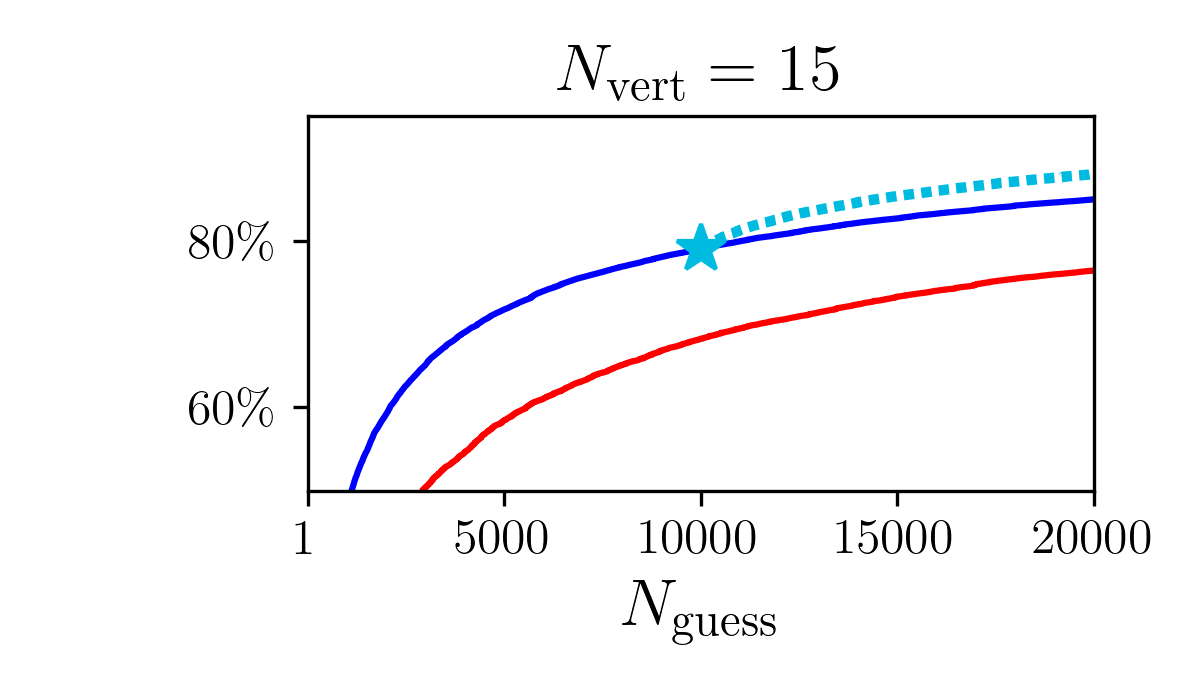}
    \end{minipage}

    % --- Legend ---
    % \vspace{0.2cm}
    \includegraphics[width=0.65\textwidth,trim={0.5cm 0.5cm 0.5cm 0.5cm},clip]{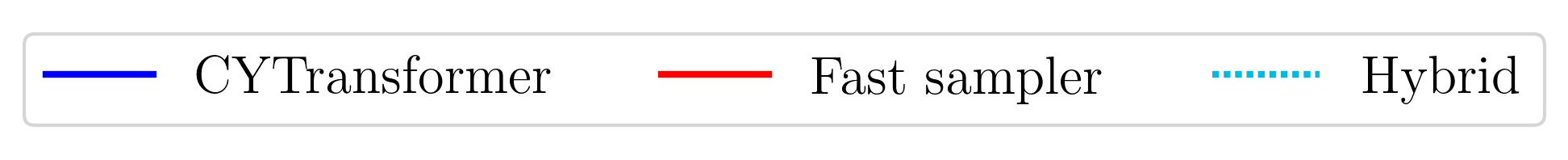}
  \caption{\textbf{Comparison between CYTransformer and the fast sampler.} Average percentage of distinct FRSTs recovered as a function of the number $N_{\rm guess}$ of inference calls, averaged over 200 test polytopes, when sampling from either CYTransformer, \texttt{CYTools}'s fast sampler, or the hybrid method (dashed, starting at the step indicated by the star) and for various $N_{\rm vert}$.
  }
  \label{fig:compare_recovery}
\end{figure*}

\subsection{Performance of trained models}
For each $N_{\rm vert}$ configuration, we select the best performing trained model and refer to it as a trained CYTransformer. The rest of this section analyzes these models in detail.

\textbf{Efficiency.} We analyzed how efficient CYTransformers are at generating FRSTs, and observed that they are competitive with pre-existing methods.
We selected $200$ test polytopes and sampled candidate FRSTs from a trained CYTransformer as well as using \texttt{CYTools}'s fast sampler algorithm. The weighted average percentage of unique FRSTs recovered with either method as a function of the number of guesses is reported  in Figure~\ref{fig:compare_recovery} (see also Figure~\ref{fig:ind_recovery} in Appendix~\ref{app:add_results} for a similar study on a per-polytope basis).

For small configurations, the FRST space itself is small (typically on the order of $10$ FRSTs per polytope for $N_{\rm vert} = 10$), allowing the fast sampler to efficiently scan the space via random perturbations. 
By contrast, CYTransformer outperforms the fast sampler for the larger, more complex polytopes with a much bigger FRST space by capturing the global distribution of FRSTs more effectively, while the fast sampler remains a local method, confined to a comparatively narrow subset of triangulations.

As CYTransformer seems to act as a fair global explorer of the FRST space, while the fast sampler serves as an efficient local scanner, we also test for $N_{\text{vert}}=13,14,15$ a hybrid strategy that combines both models' complementary strengths.
We first run CYTransformer for $N_{\rm cut}$ inference calls and collect all distinct FRSTs generated during this phase. These FRSTs form a seed pool. For the remaining $20{,}000-N_{\rm cut}$ candidate triangulations, we sample a seed uniformly at random from this pool to initialize the fast sampler.
%, which then performs a single local perturbation to generate a new candidate triangulation.
This combined method is shown to perform even better.
As transformer inference is more expensive than fast sampling, the hybrid method also allows early stopping of CYTransformer and continuation with the cheaper fast sampler, making the generation process more cost-effective.

\textbf{Representativeness.} We now want to assess how well CYTransformer’s output distribution matches the true underlying distribution of FRSTs.
As mentioned earlier, a (non-unique) height vector can be associated to every regular triangulation.
This allows us to define for each reflexive polytope and set of FRSTs associated to that polytope an \textit{FRST distribution histogram} that conveniently summarizes some of the geometric structure of this set of FRSTs. %(details are given in Appendix~\ref{app:exp_details}). 

For each test polytope, we can now compare the FRST distribution histograms associated to samples generated by either CYTransformer or \texttt{CYTools}'s fast sampler's output distribution and to the complete set of FRSTs of the polytope.
We find that the CYTransformer's histograms consistently (across the $200$ test polytopes) match the shape of the full population distribution, while the fast sampler's histograms often exhibit skewed profiles. This suggests that although both methods can eventually cover much of the FRST space, they do so differently: the CYTransformer samples in proportion to the true density of FRSTs, while the fast sampler tends to concentrate on particular regions.
Two examples are given in Figure~\ref{fig:compare_hist_main_text} (more can be found in Appendix~\ref{app:add_results}).

We make this observation quantitative by considering a more global measure of representativeness. 
For a given distribution of FRSTs and a polytope, we sample FRSTs from the distribution and consider the FRST histogram associated to this set. We treat it as a vector, and compute its cosine similarity with the similarly defined vector corresponding to the complete distribution of FRSTs. We compute this similarity score for each of our test polytopes, and plot the resulting distribution of scores as a histogram, the \textit{representativeness histogram}. The more concentrated around $1$ this histogram is, the more representative the distribution under consideration is.

We compare at the bottom of Figure~\ref{fig:compare_hist_main_text} the representativeness histograms associated to \texttt{CYTools}'s fast sampler and to CYTransformer. We see that CYTransformer  achieves higher representativeness scores with lower variance, indicating that its sampling distribution more faithfully captures the true structure of the FRST space. Similar histograms are given for $N_{\rm vert}=14,15$ in Figure~\ref{fig:compare_hist_app} of Appendix~\ref{app:add_results}.

\begin{figure}
    \centering
    \includegraphics[width=0.75\linewidth,trim={0 0 0 1.15cm},clip]{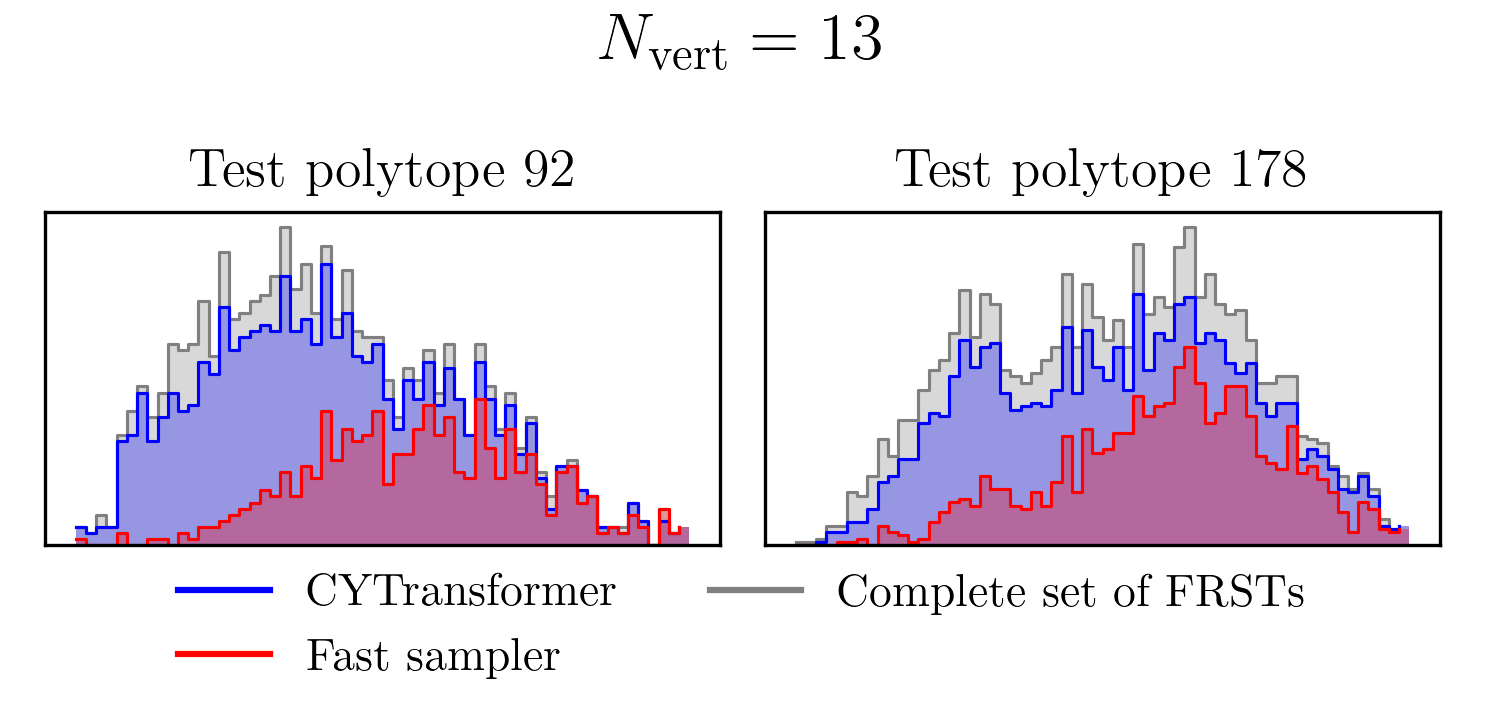}
    \includegraphics[width=0.55\textwidth]{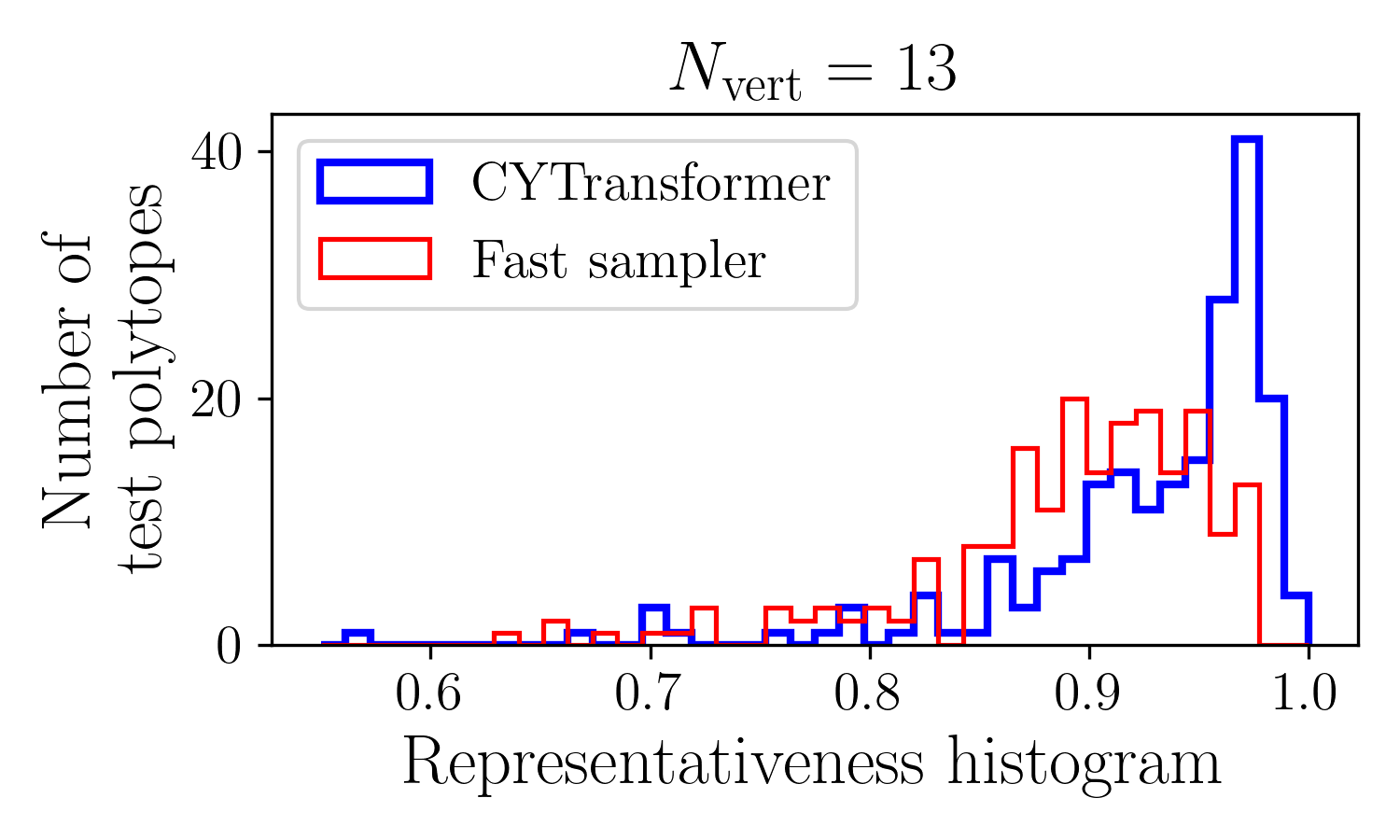}
    \caption{\textbf{Representativeness of the output distribution.} Top: FRST distribution histograms are shown for 2 test polytopes with $N_{\rm vert}=13$. The histograms are generated using the first 90\% of distinct FRSTs recovered by each method, shown alongside the full population distribution (gray).
    Bottom: representativeness histograms for CYTransformer and the fast sampler over $200$ test polytopes with $N_{\rm vert}=13$ vertices.
}
    \label{fig:compare_hist_main_text}
\end{figure}

\subsection{Self-improvement}

As observed in section~\ref{subsec:training_dynamics} model performance is positively correlated with the training set size. This motivates our exploration of the self-improvement strategy described in section~\ref{subsec:selfim_method}, where we start with a small initial training set and alternate between using the model to generate new training data and retraining it on the augmented training set.
Since this work focuses on smaller $N_{\rm vert}$ for which large datasets are available, our self-improvement experiments serve primarily as a proof of concept for the method.

\begin{figure}[htbp]
  \centering

  \includegraphics[height=3.8cm,trim={0cm 0 0 0},clip]{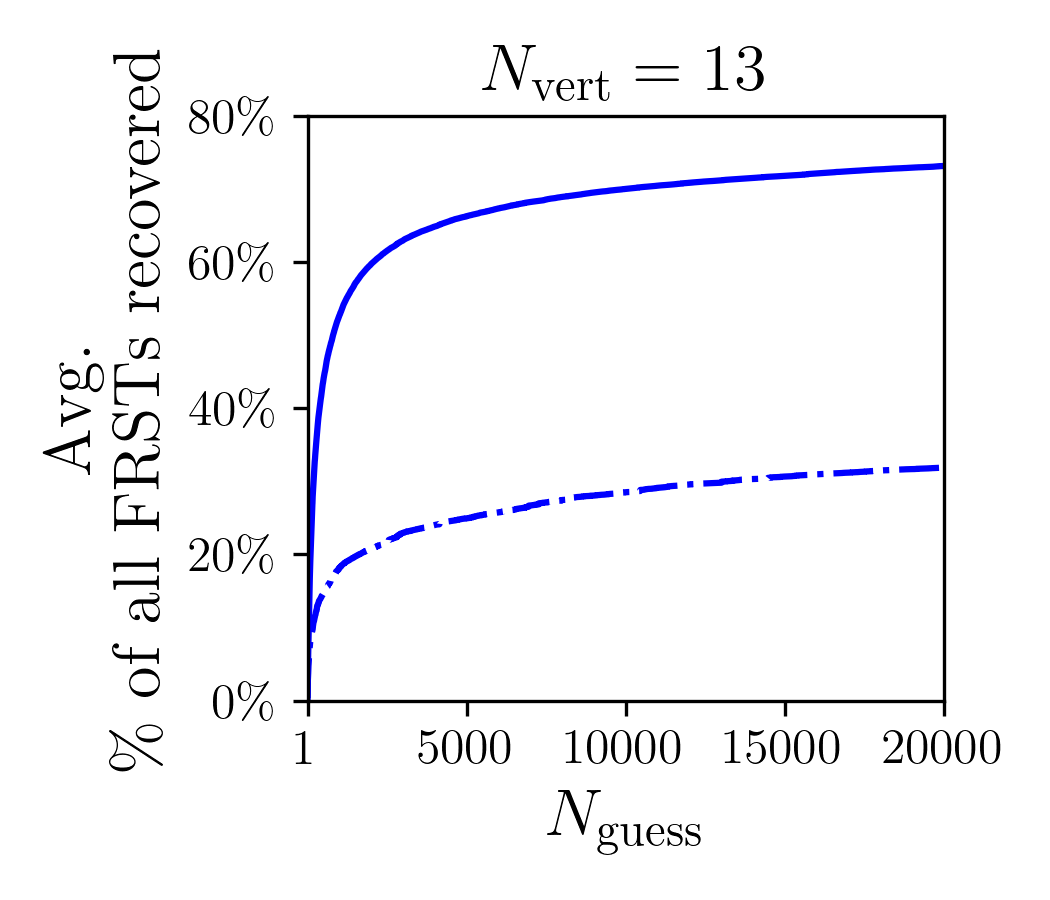}
  \includegraphics[height=3.8cm,trim={1.6cm 0 0 0},clip]{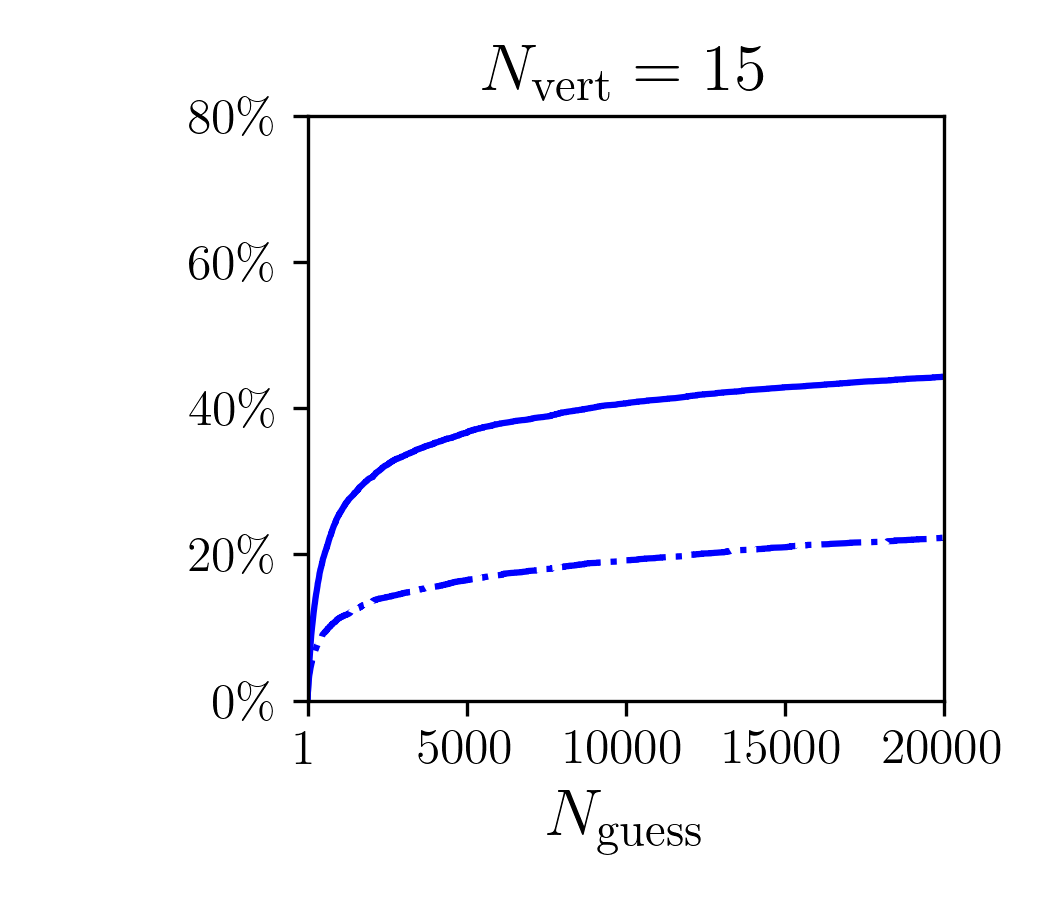}\\
  \includegraphics[width=0.5\textwidth,trim={0.5cm 0.5cm 0.5cm 0.5cm},clip]{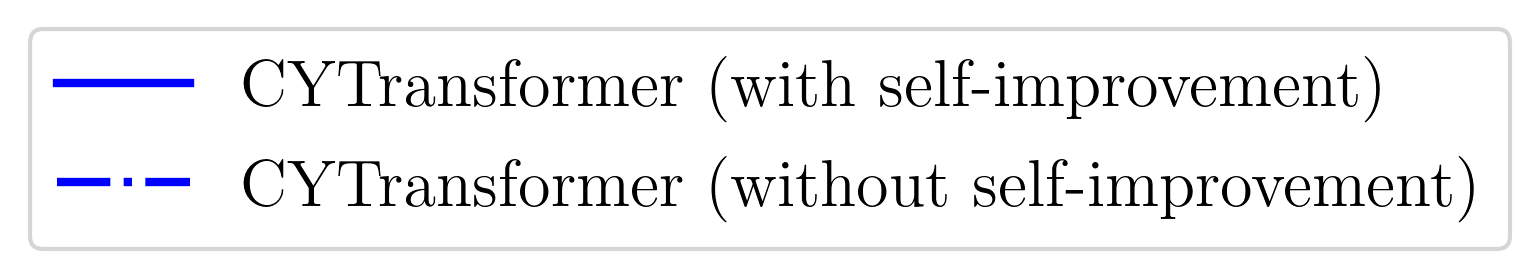}
  \caption{\textbf{Performance before and after self-improvement.} Average percentage of distinct FRSTs recovered as a function of the number $N_{\rm guess}$ of inference calls when sampling from a model trained on a small training set of at most $5$ triangulations per polytope with (solid) and without (dash-dotted) self-improvement for $N_{\rm vert} =13, 15$. }
  \label{fig:self_improvement_recovery_curves_main_text}
\end{figure}

To assess the effectiveness of this approach, we compare self-improved models to baselines trained (to convergence) solely on the same small initial datasets. Figure~\ref{fig:self_improvement_recovery_curves_main_text} shows that self-improved CYTransformer models achieve up to $229$\% higher average FRSTs recovered on the test set than the baselines. %(see Figure~\ref{fig:self_improvement_recovery_curves_app} in Appendix~\ref{app:add_results} for other $N_{\rm vert}$). 
This demonstrates that self-improved models can greatly surpass those trained with straightforward supervised training in data-scarce regimes. However, they still fall short of the performance achieved by models trained on much larger datasets (cf. Figure~\ref{fig:compare_recovery}) in the most complex cases. This gap is primarily due to the self-improved models failing on a fraction of test polytopes, suggesting that for some polytope geometries, there exists a minimal threshold of initial data required for self-improvement to be effective.
We also report in Appendix~\ref{app:add_results} FRST distribution histograms associated to the self-improved model's outputs, which suggest that similarly to the models considered above, the self-improved model learns an unbiased representation of the FRST space for those polytope geometries.

\section{Discussion}\label{sec:dis}

In this work, we applied a transformer architecture to generate FRSTs, special triangulations of four-dimensional polytopes that satisfy highly nontrivial conditions. The method demonstrates promising performance across the range of polytope sizes considered, particularly in its ability to efficiently and unbiasedly sample the space of triangulations. Nevertheless, several avenues for future research remain, such as enabling knowledge transfer from a given polytope size to a larger one, extending our method to other distributions of triangulations of interest, or optimizing our pipeline to scale to larger polytopes.

\section{Acknowledgments}\label{sec:ack}

The work of G.S. and J.H.T.Y.
is supported by the U.S. Department of Energy, Office of Science, Office of High Energy
Physics under Award Numbers DE-SC-0023719 and DE-SC-0017647. F.C. was affiliated
with Meta FAIR during the period when the main part of this research was conducted.

\begin{appendix}

\section{Additional experimental results}\label{app:add_results}

% \refine{Improve section}

\begin{figure*}[t]
    \centering
    \includegraphics[height=10cm]{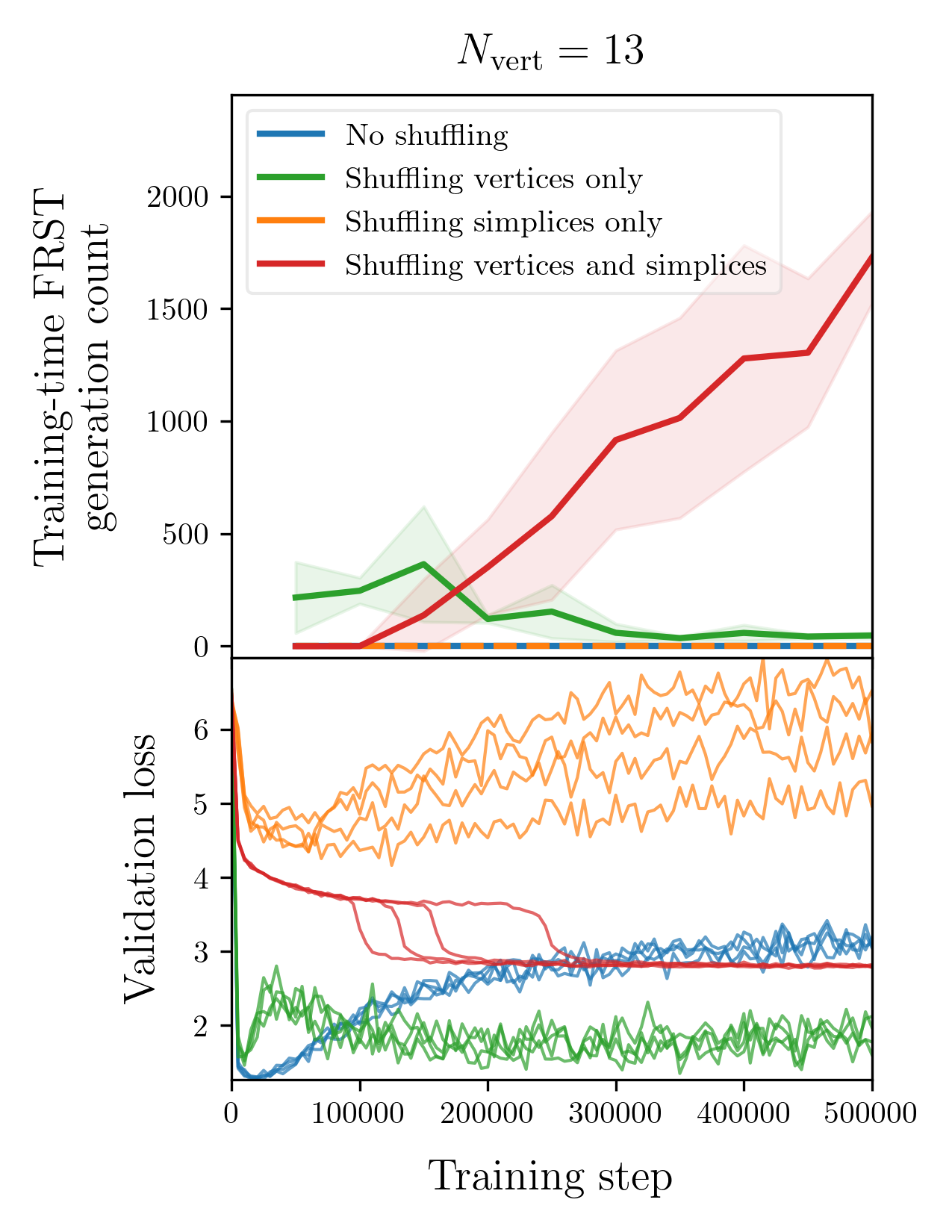}
  \caption{\textbf{Data augmentation ablations} Training time FRST generation curves (left, average and std.) and test loss (right) for $N_{\rm vert} = 13$, a training set of $7500$ polytopes with up to $200$ triangulations for each polytope, and all four combinations of data augmentation procedures: with and without permutation of the vertices within a polytope, and with or without permutation of the simplices within the sequences that represent FRSTs. Each configuration is run with $4$ distinct random seeds. }
  \label{fig:data_augmentation}
\end{figure*}

\begin{figure}[htbp]
  \centering
  \makebox[\textwidth][c]{%
  \includegraphics[height=5.25cm]{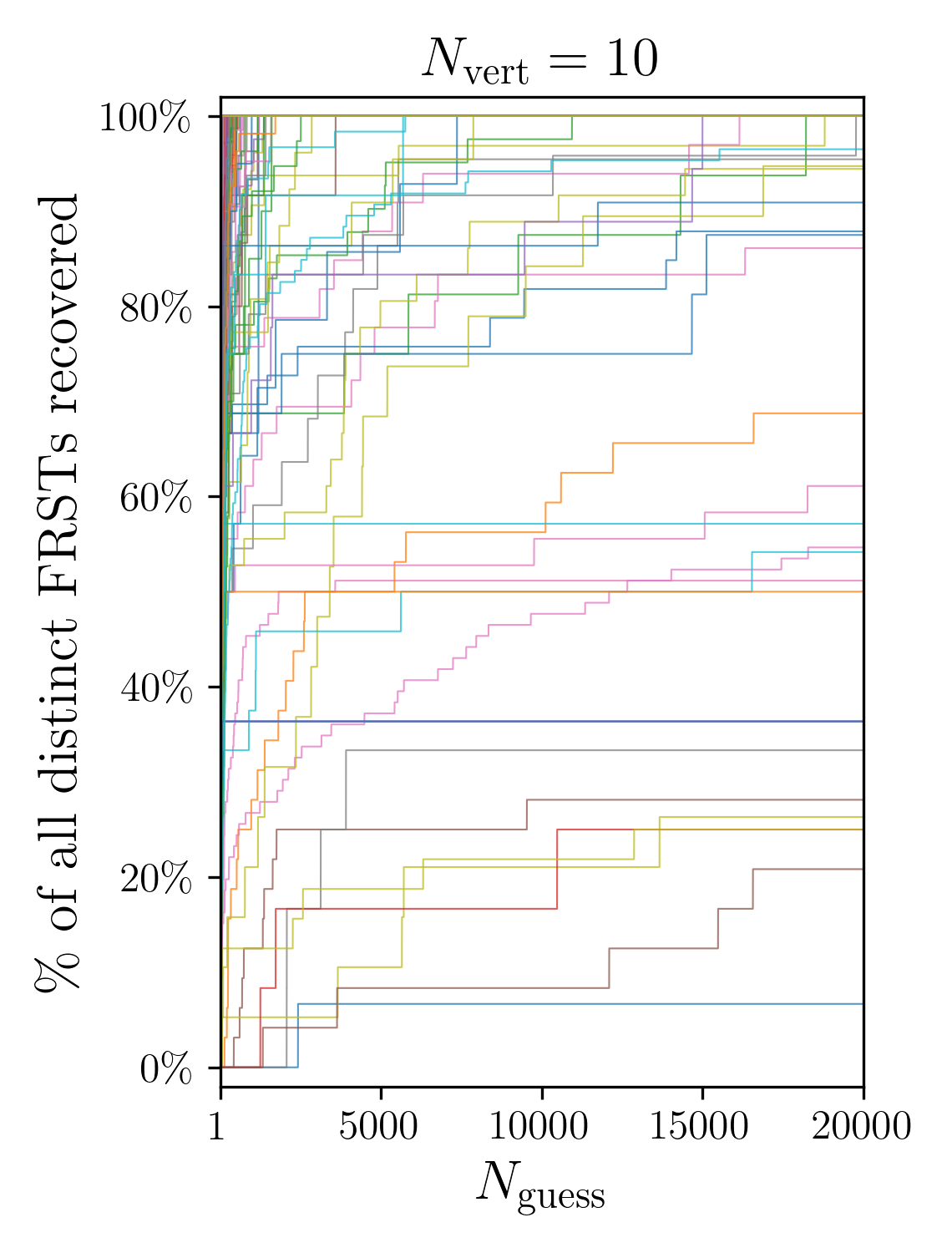}
  \includegraphics[height=5.25cm,trim={1.00cm 0 0 0},clip]{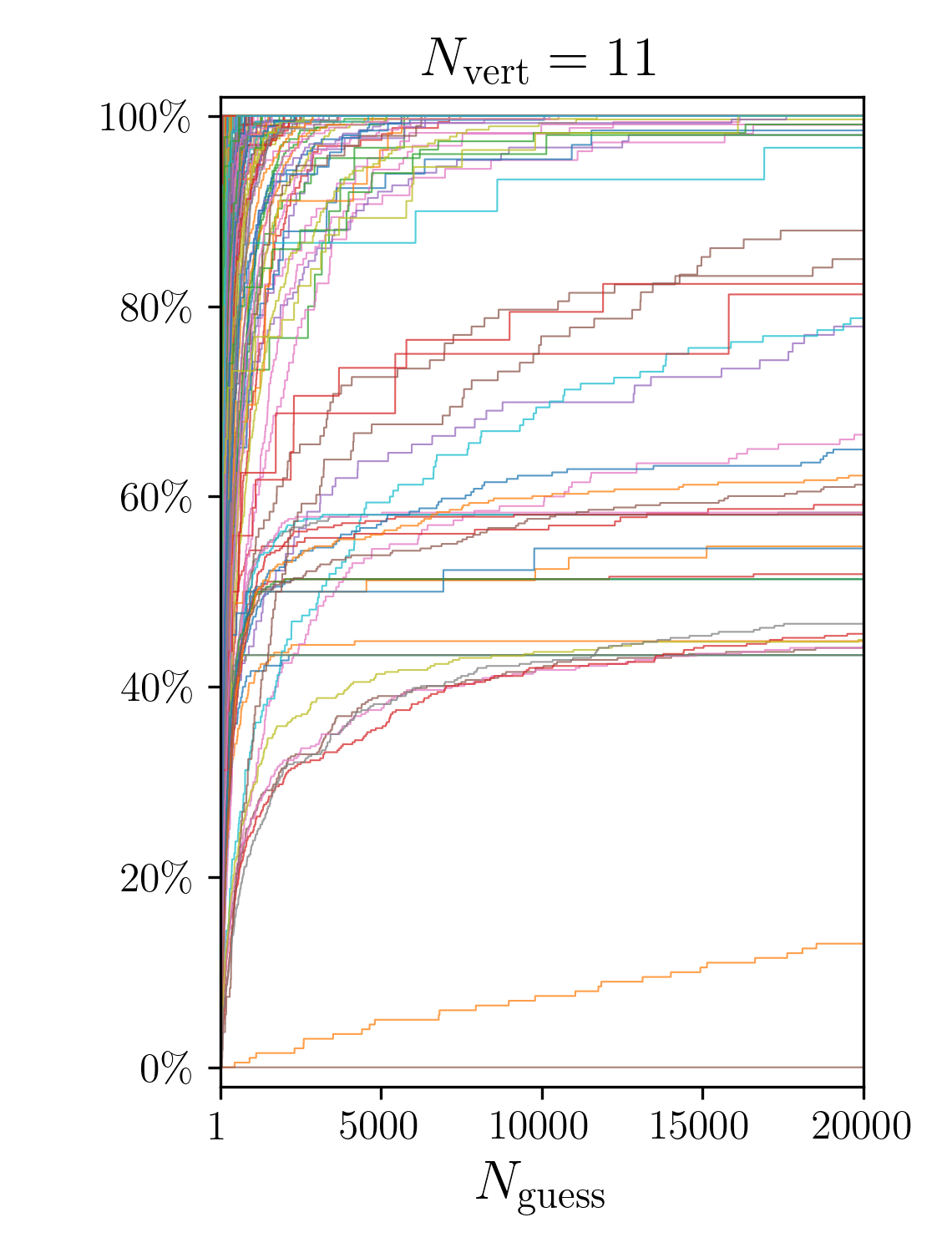}
  \includegraphics[height=5.25cm,trim={1.00cm 0 0 0},clip]{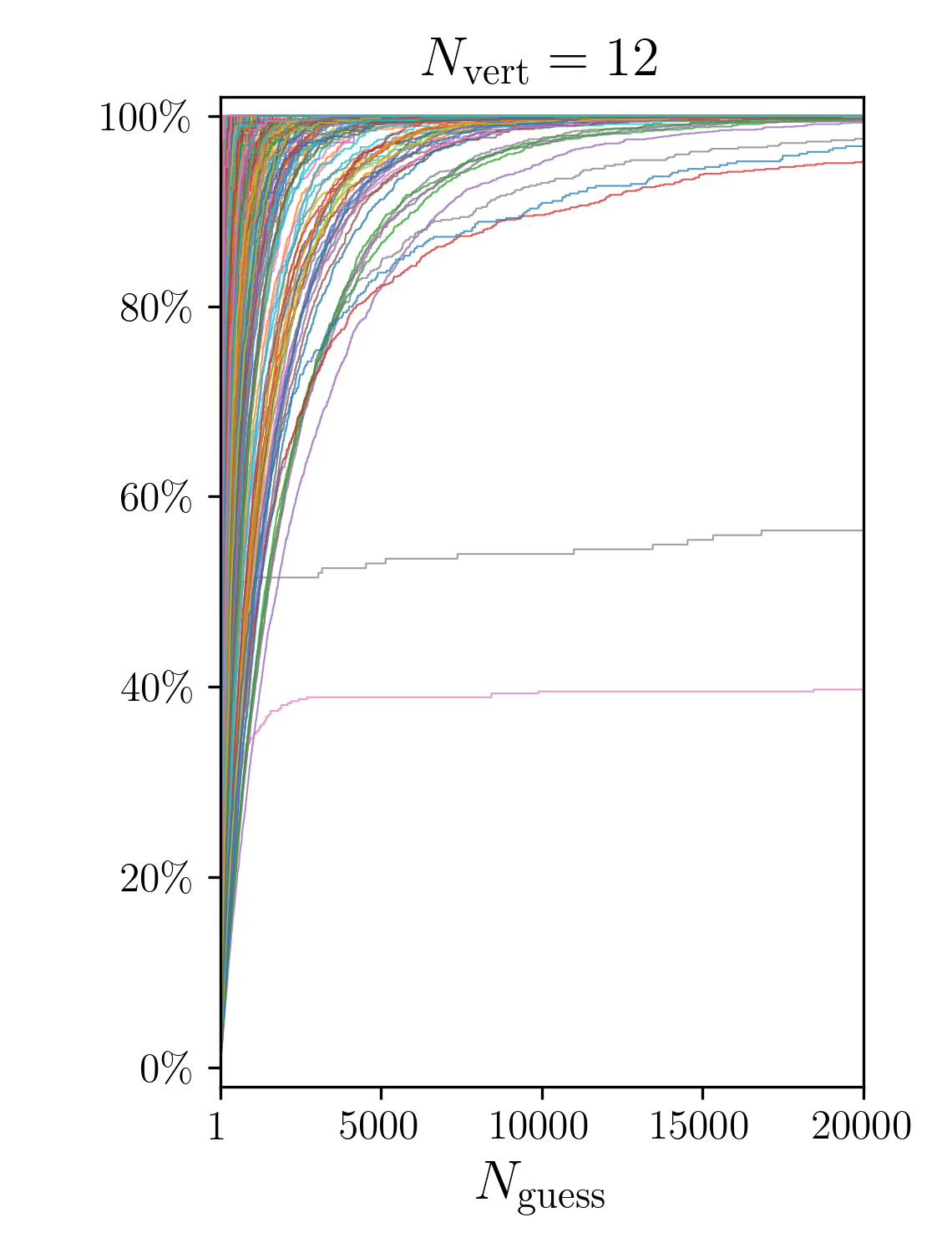}}
  \makebox[\textwidth][c]{%
  \includegraphics[height=5.25cm]{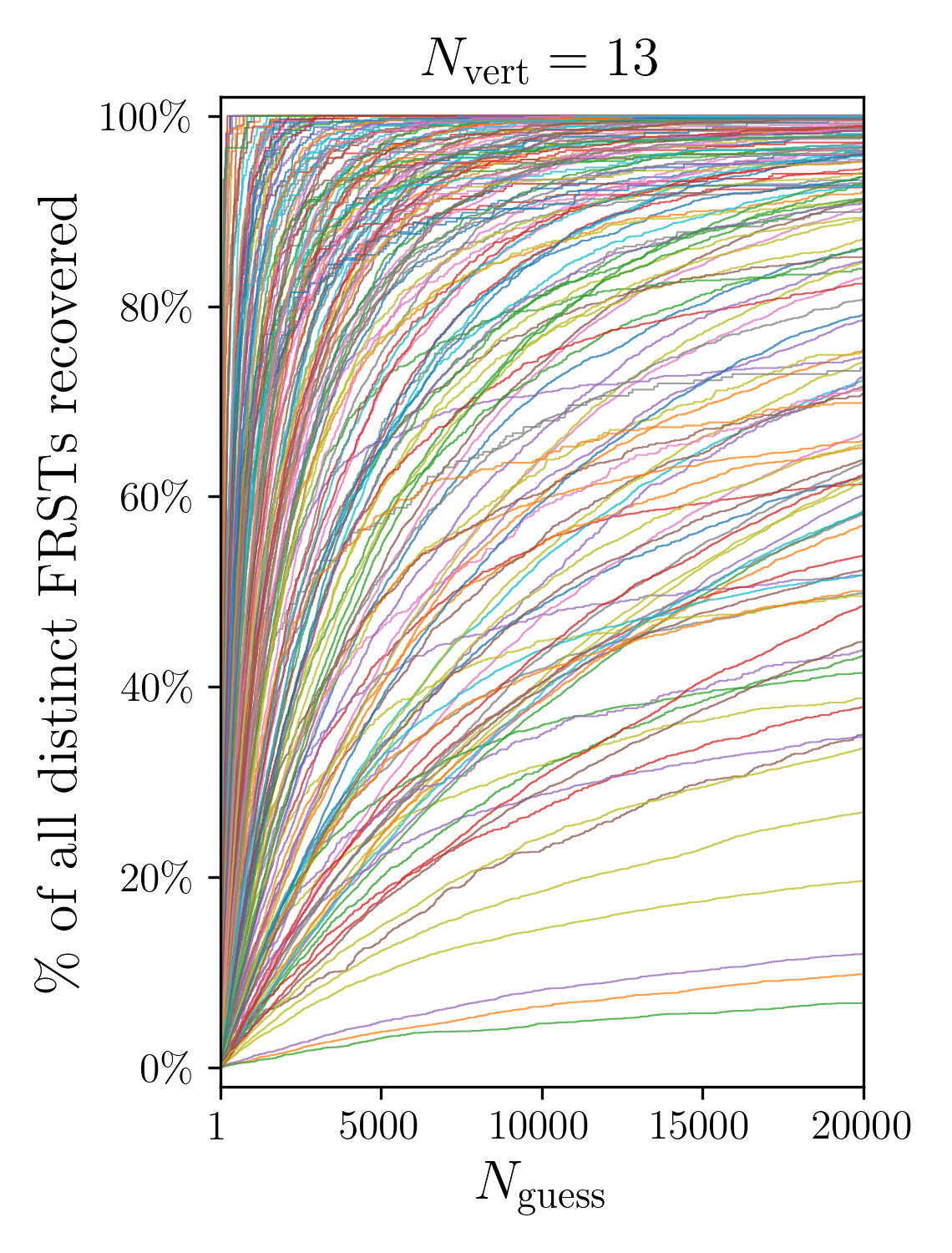}
  \includegraphics[height=5.25cm,trim={1.00cm 0 0 0},clip]{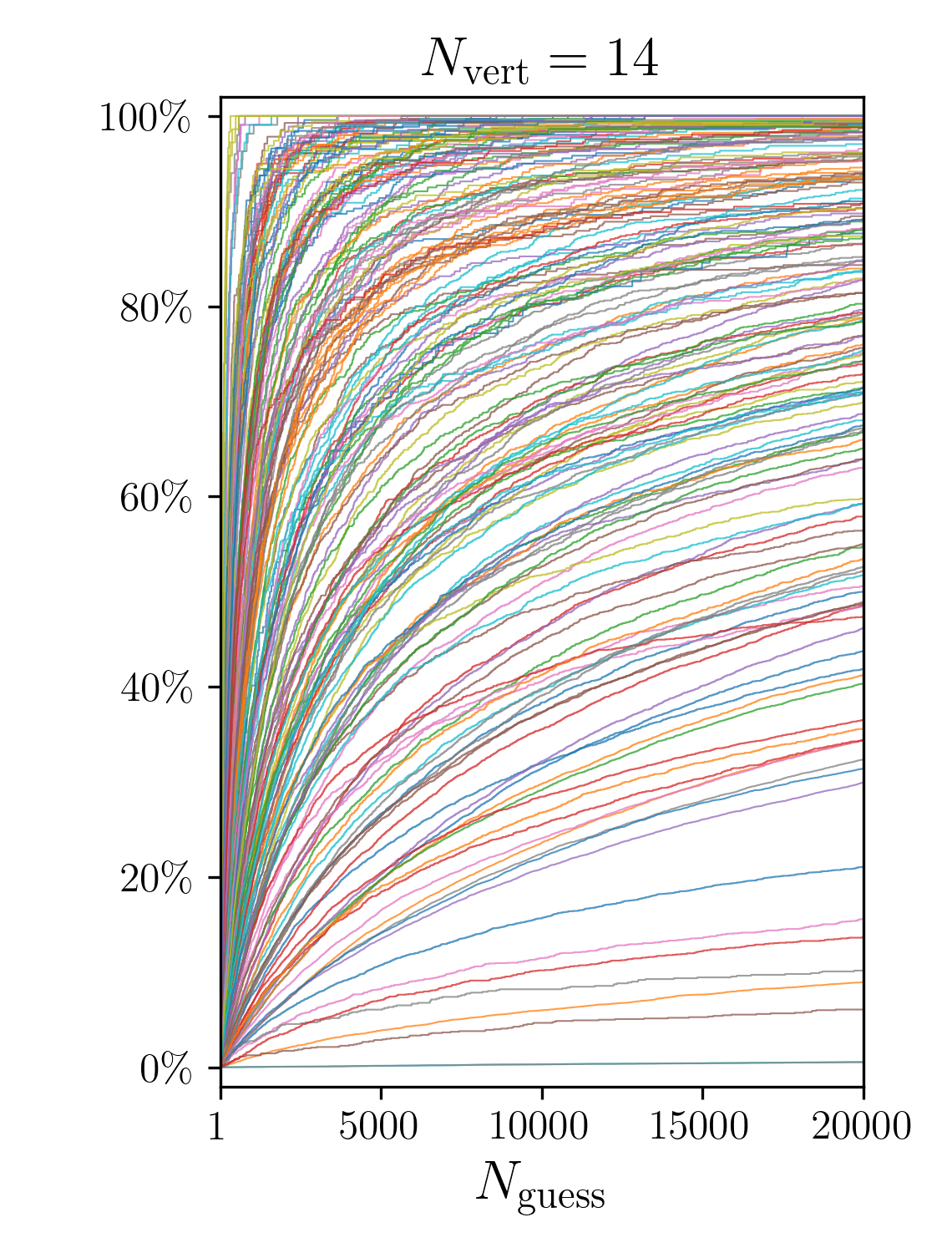}
  \includegraphics[height=5.25cm,trim={1.00cm 0 0 0},clip]{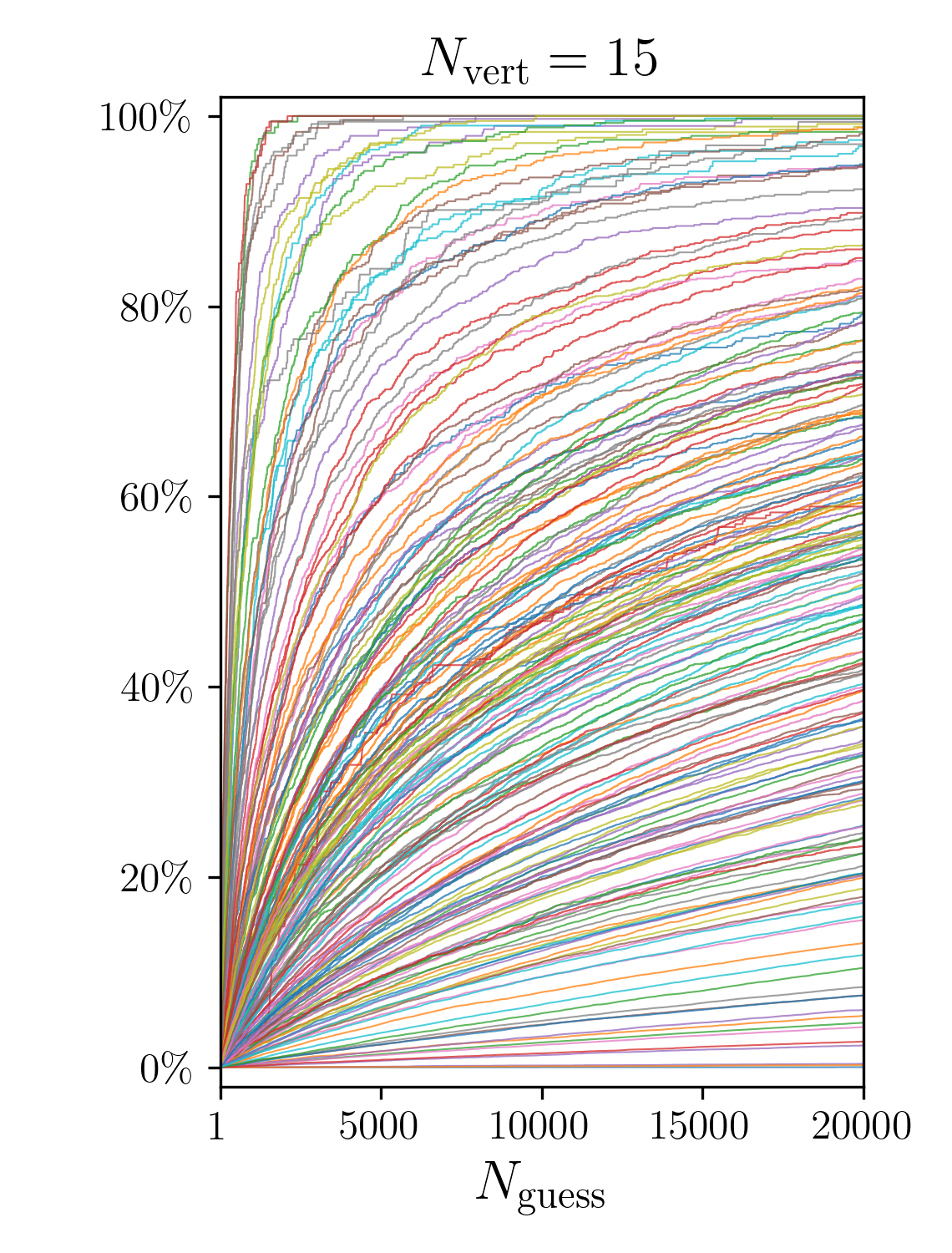}}
  \caption{  \textbf{Per-polytope FRST recovery curves.} Each panel shows the percentage of all distinct FRSTs recovered as a function of inference calls $N_{\rm guess}$, plotted individually for test polytopes within a fixed $N_{\rm vert}$ set. While most polytopes exhibit rapid and high recovery, a noticeable subset show slow or limited recovery, highlighting the influence of polytope geometry on model performance. Some curves plateau early or rise slowly, indicating cases where CYTransformer struggles to learn the full FRST space efficiently. Common in the $N_{\rm vert}=15$ case, smooth and gently sloping curves correspond to polytopes with especially large FRST spaces, for which higher sampling budgets are necessary to achieve meaningful recovery.}
  \label{fig:ind_recovery}
\end{figure}

\begin{figure}
    \centering
    \makebox[\textwidth][c]{%
    \includegraphics[width=1\linewidth]{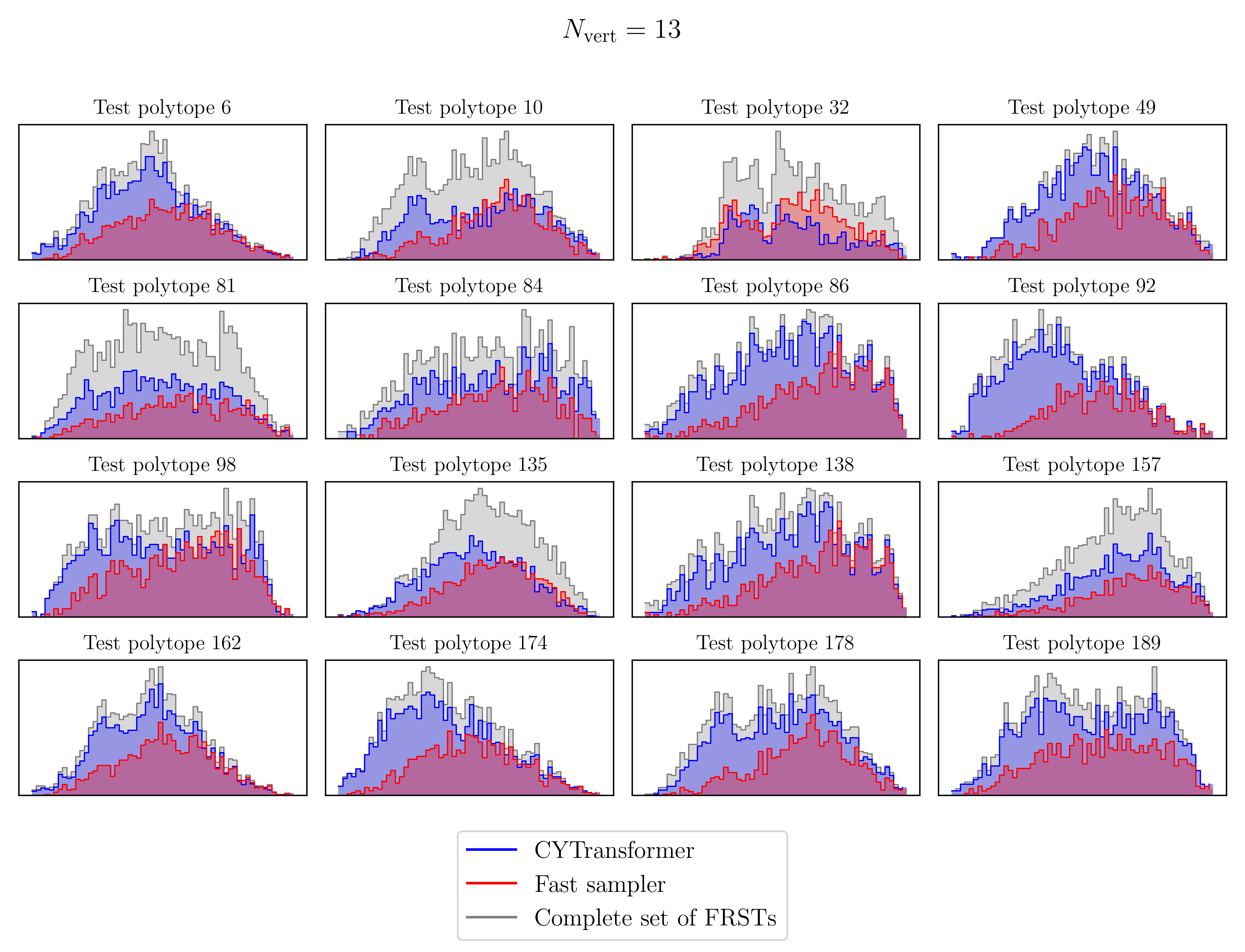}
    }
    \caption{\textbf{Comparison of sampling distributions between CYTransformer and the fast sampler.} Height-space FRST distributions are shown for 16 test polytopes with $N_{\rm vert}=13$. These examples are selected to highlight contrasting sampling behaviors. For each polytope, we compare CYTransformer (blue) and the fast sampler (red) using the first 90\% of distinct FRSTs recovered by each method, shown alongside the full population distribution (gray). The 90\% threshold is used purely for visualization clarity. CYTransformer consistently matches the shape of the population histogram more closely, indicating more representative and unbiased sampling across the FRST space. In contrast, the fast sampler often exhibits skewed distributions, reflecting sampling bias or overconcentration in specific regions.
}
    \label{fig:compare_hist_app}
\end{figure}

\begin{figure}
  \centering
  \makebox[\textwidth][c]{%
  \includegraphics[width=0.5\textwidth]{plots/rep_12.png}
  \includegraphics[width=0.5\textwidth]{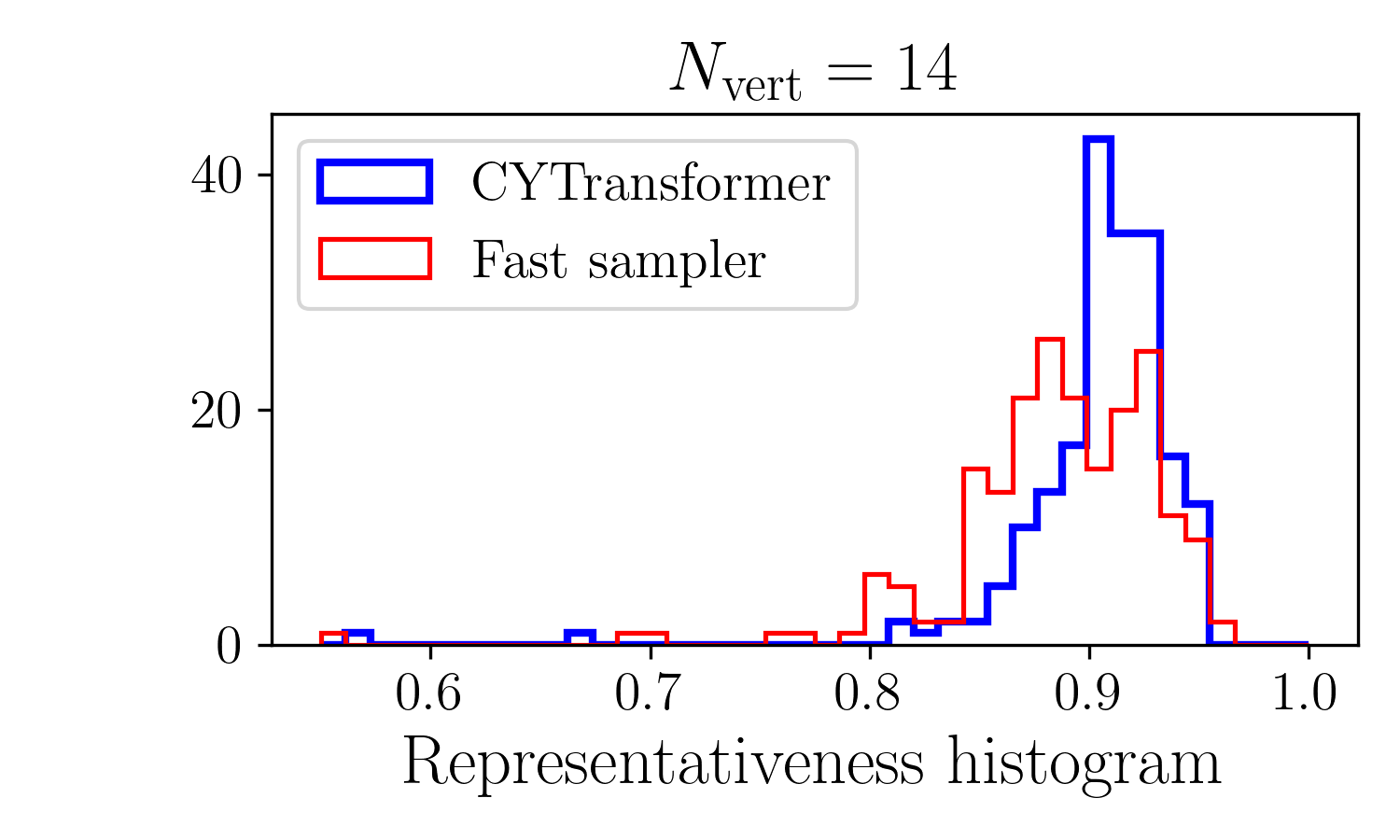}}
  \makebox[\textwidth][c]{%
  \includegraphics[width=0.5\textwidth]{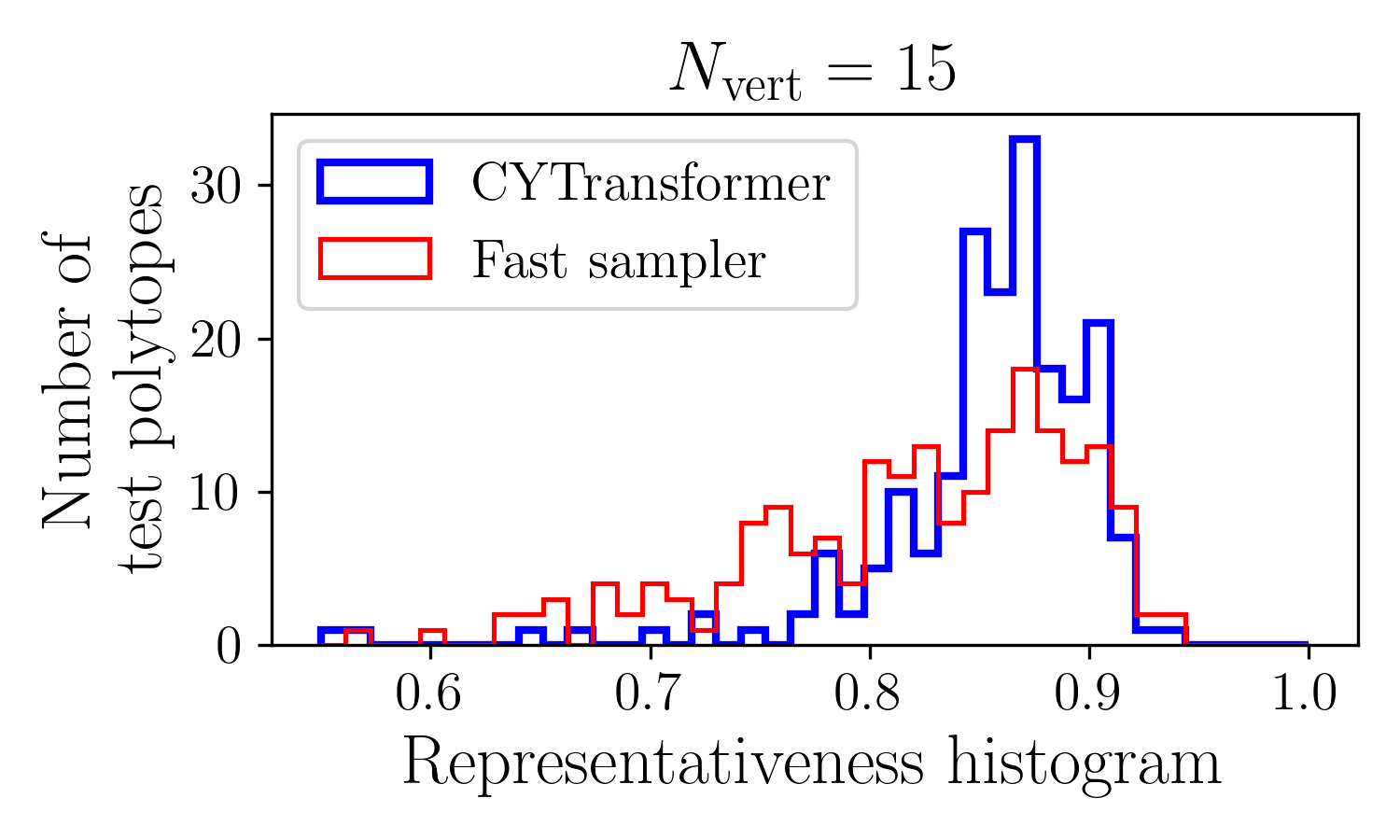}}
  \caption{\textbf{Representativeness histograms.} For each of the $200$ test polytopes,  we compute the cosine similarity between the method's FRST histogram and that of the full population distribution of the same polytope. Across all $N_{\rm vert}$ cases shown, the distributions of scores for CYTransformer (blue) are sharply peaked near unity with low variance, indicating that it consistently and unbiasedly produces samples that reflect the true shape of the FRST distribution. In contrast, the fast sampler (red) exhibits broader and flatter score distributions, confirming that it introduces a stronger sampling bias and tends to favor certain regions of the FRST space over others.}
  \label{fig:compare_rep}
\end{figure}

We present additional experimental results:
\begin{itemize}
    % \item We report FRST generation count as a function of training step for various $N_{\rm vert}$ and training sets sizes in Figure~\ref{fig:training_dynamics_app}.
    \item We conduct ablations on data augmentation techniques in Figure~\ref{fig:data_augmentation}.
    % \item We conduct ablations on model size in Figure~\ref{fig:model_size}.
    \item We study on a per-polytope basis the FRST recovery rate in Figure~\ref{fig:ind_recovery}.
    \item We compare the FRST histograms associated to CYTransformer and to the fast sampler for 16 polytopes in Figure~\ref{fig:compare_hist_app}, and report the full representativeness histograms in Figure~\ref{fig:compare_rep}.
    % \item Finally, we study the self-improvement more in depth by providing FRST recovery curves for additional $N_{\rm vert}$ in Figure~\ref{fig:self_improvement_recovery_curves_app }, by reporting the growth of the partially self-generated training set in Figure~\ref{fig:self_improvement_data }, and by studying the representativeness of the resulting distribution in Figure~\ref{fig:self_improvement_histogram}.
\end{itemize}

% \subsection{Training dynamics}

% \begin{figure*}[t]
%     \centering
%   \includegraphics[height=4cm]{plots/frst_plot_9p1_app.png}
%   \includegraphics[height=4cm,trim={1.0cm 0 0 0},clip]{plots/frst_plot_10p1_app.png}
%   \includegraphics[height=4cm,trim={1.0cm 0 0 0},clip]{plots/frst_plot_11p1_app.png}
%   \includegraphics[height=4cm]{plots/frst_plot_12p1_app.png}
%   \includegraphics[height=4cm,trim={1.0cm 0 0 0},clip]{plots/frst_plot_13p1_app.png}
%   \includegraphics[height=4cm,trim={1.0cm 0 0 0},clip]{plots/frst_plot_14p1_app.png}
%   \caption{ \textbf{CYTransformer training-time FRST generation curves.} Distinct FRSTs generated during training, measured across $1{,}600$ (for $N_{\rm vert})=10,11,12,13$) or $6{,}400$ (for $14,15$) candidate triangulations, as a function of training step.  Each curve within a plot corresponds to a model trained on a different dataset size $(N_{\rm polys}, N_{\rm triangs})$, where $N_{\rm polys}$ is the total number of polytopes and $N_{\rm triangs}$ is the maximum number of FRSTs per polytope (with "all" being all available FRSTs). 
%  }
%   \label{fig:training_dynamics_app}
% \end{figure*}

% \subsection{Ablations on the data augmentation procedure}

% \charles{
% The average generation curves mask the variability that arises from polytope-specific geometry. To uncover this, we turn to individual \hyperlink{met:frstreccurve}{recovery curves}, where each curve represents the number of distinct FRSTs generated, normalized by the total number admitted by the polytope. These plots, shown in figure~\ref{fig:ind_recovery}, illustrate that generation efficiency varies across polytopes, highlighting the impact of polytope geometry on model performance.\footnote{We leave the investigation of how polytope geometry affects learning to future work.} While most test polytopes exhibit recovery curves that quickly reach a high saturation level, a noticeable number show limited or slower recovery. For example, in the $(6,10+1)$ panel, the brown horizontal line at the bottom represents a case where the model fails to recover any FRSTs at all. Just above it, the orange curve shows a steady yet slow increase, suggesting that CYTransformer does learn something, though not very effectively. These rare but poorly performing cases may be attributed to the model’s difficulty in generalizing to uncommon polytope geometries or the inherent complexity of the corresponding FRST spaces. Finally, a smooth and gently sloping curve often corresponds to a polytope that admits a large number of FRSTs; such polytopes are especially prevalent in the $(10,14+1)$ case (curves near the bottom of the plot), where significantly more inference calls are needed to achieve a meaningful recovery rate.
% }

% \clearpage

% \subsection{Representativeness}

\end{appendix}

\clearpage
\bibliographystyle{ws-rv-van}
\bibliography{biblio}

%\blankpage
%\printindex[aindx]                 % to print author index
%\printindex                        % to print subject index

\end{document}